\documentclass[journal,onecolumn]{IEEEtran}
\usepackage{hyphenat}
\usepackage{amsmath,amssymb,amsthm,mathtools}
\usepackage{booktabs}
\usepackage{array}
\usepackage{graphicx}
\usepackage{cite}
\usepackage{microtype}
\usepackage{algorithm}
\usepackage{algpseudocode}
\usepackage{bm}
\usepackage{xcolor}
\newtheorem{theorem}{Theorem}
\newtheorem{lemma}[theorem]{Lemma}

\newtheorem{corollary}[theorem]{Corollary}
\theoremstyle{definition}
\newtheorem{definition}[theorem]{Definition}
\theoremstyle{remark}
\newtheorem{remark}[theorem]{Remark}
\newcommand{\D}{\mathcal D}
\newcommand{\C}{\mathcal C}
\newcommand{\A}{\mathcal A}
\newcommand{\B}{\mathcal B}
\newcommand{\E}{\mathbb E}
\renewcommand{\S}{\mathcal S}
\newcommand{\Prob}{\mathbb P}
\newcommand{\del}{\mathrm{del}}

\newcommand{\Sub}
{\operatorname{Sub}}
\newcommand{\Diff}{\operatorname{Diff}}
\newcommand{\Int}{\operatorname{Int}}
\newcommand{\Var}{\operatorname{Var}}
\newcommand{\ind}{\mathbf 1}
\newcommand{\eps}{\varepsilon}

\theoremstyle{plain}
\newtheorem*{lemhinv}{Lemma \ref{lem:hinv}}

\title{Minimum Synthesis Time for Single-Deletion Correcting Codes of Prescribed Size}

\author{Yuling~Li, Wenjun~Yu, Gennian~Ge
\thanks{This research was supported by the National Key Research and Development Program of China under Grant 2025YFC3409900, the National Natural Science Foundation of China under Grant 12231014, and Beijing Scholars Program.}
\thanks{
Yuling Li and Gennian Ge are with the School of Mathematical Sciences, Capital Normal University, Beijing 100048, China (e-mail: 2240501022@cnu.edu.cn;  gnge@zju.edu.cn).
}
\thanks{Wenjun Yu is with the Institute of Mathematics and Interdisciplinary Sciences, Xidian University, Xi’an 710126, China (e-mail: yuwenjun@xidian.edu.cn).}
}

\begin{document}
\maketitle

\begin{abstract}
DNA is a promising medium for data storage, but the high
cost of DNA synthesis remains a major obstacle to its
widespread deployment. In the reference-based synthesis
model, synthesis time is measured by the length of the
reference sequence. Previous studies have analyzed code
sizes under synthesis-time constraints and constructed
cost-constrained codes, including codes that correct deletions.
In this paper, we study the minimum synthesis time for codes
of prescribed size, considering both unrestricted codes and
single-deletion correcting codes. We establish finite-length
lower bounds valid for arbitrary reference sequences and
upper bounds using existing $q$-ary differential VT codes.
For fixed $q\geq3$, $A>0$, and $R\in(0,1)$, we determine
the minimum synthesis time up to an $O(1)$ term when the
code size is $M_n=\lfloor Aq^{Rn}\rfloor$.
The corresponding redundancy is
$(1-R)n\log_2q-\log_2A+o(1)$ bits.
In particular, for these code sizes, the minimum synthesis
time for single-deletion correcting codes exceeds that for
unrestricted codes by a term proportional to $\ln n$,
and we determine its coefficient.
We also derive upper and lower bounds for code sizes
$\lfloor q^n/(nq^{s_n})\rfloor$, where $s_n\to\infty$
and $s_n=o(n)$.
These sizes correspond to redundancy
$\log_2 n+s_n\log_2q+o(1)$ bits.
For code sizes $\lfloor q^n/(nq^c)\rfloor$ with fixed constant $c$,
the redundancy is $\log_2 n+c\log_2q+o(1)$ bits.
We give conditions under which single-deletion correcting
codes exist or do not exist for all sufficiently large $n$,
together with bounds on the minimum synthesis time.
These results show how code size and the single-deletion
correcting constraint affect synthesis time and provide
theoretical guidance for the construction of cost-constrained codes.
\end{abstract}


\section{Introduction}
DNA is a promising medium for data storage because of its
high storage density and long-term stability.
In a DNA storage system, digital information is encoded
into nucleotide sequences, written into DNA molecules
through synthesis, and retrieved through sequencing.
The cost and speed of synthesis remain important obstacles
to large-scale DNA storage \cite{Yu2024Synthesis}.
Errors introduced during synthesis, storage, and sequencing
also require suitable coding methods.
Thus, efficient DNA storage calls for codes that reduce
synthesis time while providing error correction.

In the conventional synthesis model, multiple strands are
grown in parallel according to a reference sequence.
During each cycle, the machine processes the next symbol
of the reference and appends the corresponding nucleotide
to a selected subset of the strands.
Each selected strand receives one nucleotide, while the
other strands remain unchanged.
A strand can therefore be synthesized if and only if it
is a subsequence of the reference, and the synthesis time
is the length of the reference prefix used.

Under this mechanism, one question is how to maximize
the code size within a prescribed number of synthesis
cycles when the reference sequence is fixed.
In 2020, Lenz et al. \cite{Lenz2020} showed that adding
redundancy can reduce synthesis time and related the
maximum information per cycle for a periodic reference
to the capacity of a cost-constrained channel.
Precise estimates for the number of sequences were developed
in \cite{LenzMelczer2025}, while constructions and analyses
of the tradeoff between redundancy and average or maximum
synthesis time were studied in
\cite{Immink2024Synthesis,ImminkMaxentropic2024}.
A different question arises when the strands to be
synthesized are given and the reference can be chosen.
Makarychev et al. \cite{Makarychev2022} studied how to
partition a given pool of random strands into batches
and obtained upper and lower bounds on the total reference
length required for the batches.
For random strands with a bound on homopolymer length,
Elishco and Huleihel \cite{ElishcoHuleihel2023} proved
the asymptotic optimality of an alternating reference
under their Markov model.

Error correction and sequence constraints have also been
considered under the same synthesis mechanism.
For deletion and insertion errors, Chrisnata et al.
\cite{Chrisnata2023} constructed quaternary codes with
bounded synthesis time under an alternating reference
and provided polynomial-time encoders.
Codes with a prescribed minimum Hamming distance were
studied through Gilbert--Varshamov bounds in
\cite{Goyal2023}.
A different source of deletions is a synthesis defect:
when a synthesis cycle fails, the strands scheduled to
receive a nucleotide in that cycle lose the corresponding
symbol. These deletions are therefore associated with
the same failed cycle across the affected strands.
Lu et al. \cite{Lu2024Defects} constructed codes correcting
synthesis defects when the defective cycles are known
and when they are unknown.
For biological constraints, Nguyen et al.
\cite{NguyenImmink2024} combined a synthesis-time constraint
with single-deletion or single-insertion correction and
a runlength constraint.
Wang et al. \cite{Wang2025Capacity} studied the number
of sequences satisfying synthesis-time, runlength,
and GC-content constraints and the corresponding
channel capacities.
Liu and Duman \cite{LiuDuman2025} developed encoding and
decoding methods under these combined constraints,
including constructions that correct a single insertion
or deletion.

There are also related studies under different synthesis
mechanisms.
In shortmer synthesis, each cycle appends a specified
short DNA strand, rather than a single nucleotide,
to selected strands.
Abu-Sini et al. \cite{AbuSini2025} studied
the choice of shortmers and reference sequences,
developed an algorithm for minimizing the number of
cycles needed to synthesize a given set of strands,
and related information rates to the capacities of
non-deterministic costly constrained graphs.
Moav et al. \cite{Moav2025} considered complex DNA
synthesis, where each strand may receive one nucleotide
chosen from a prescribed set in each cycle.
They analyzed the maximum information per cycle and
developed algorithms for finding optimal references for
given strands.

In this paper, we consider the single-nucleotide synthesis
model and study the minimum synthesis time for
a prescribed code size over a general alphabet $\Sigma_q$.
We study the minimum synthesis time required to synthesize
a code $\C\subseteq\Sigma_q^n$ with $\lvert\C\rvert=M_n$.
The corresponding minimum times for unrestricted codes
and single-deletion correcting codes are denoted by
$\tau_q(n,M_n)$ and $\tau_q^{\del}(n,M_n)$, respectively.
We first establish implicit bounds on the minimum synthesis
time. The lower bounds hold for arbitrary reference sequences,
while the upper bounds follow from codes synthesized by
alternating references. We then use precise probability
estimates to obtain explicit asymptotic expressions for these
bounds, including terms beyond the leading linear term.
For fixed $q\geq3$, our main results cover the following
three cases.

\begin{enumerate}
\item[(i)]
For $M_n=\lfloor Aq^{Rn}\rfloor$, where $A>0$ and
$R\in(0,1)$ are fixed, the redundancy grows linearly
with $n$.
We determine both $\tau_q(n,M_n)$ and
$\tau_q^{\del}(n,M_n)$ up to an $O(1)$ term,
thereby identifying their linear and logarithmic terms.
In particular, we determine the coefficient of the
additional $\ln n$ term caused by single-deletion
correction.
This gives a precise measure of the synthesis-time cost
of error correction at a prescribed limiting code rate.

\item[(ii)]
For $M_n=\lfloor q^n/(nq^{s_n})\rfloor$, where
$s_n\to\infty$ and $s_n=o(n)$, the redundancy is
$\log_2 n+s_n\log_2q+o(1)$ bits.
We obtain upper and lower bounds with the same leading
terms and explicitly determine their gap, which is of
order $\sqrt{n/s_n}$.
Under the stronger condition $s_n=o(\sqrt n)$,
the bounds can be expressed directly in terms of $s_n$.
These estimates describe how a sublinear excess redundancy
above $\log_2 n$ affects the minimum synthesis time.

\item[(iii)]
For $M_{n,c}=\lfloor q^n/(nq^c)\rfloor$ with fixed
$c\in\mathbb R$, the redundancy is
$\log_2 n+c\log_2q+o(1)$ bits.
We prove infeasibility when $c<\log_q(q-1)$ and
establish feasibility when $c\geq1$ for all sufficiently
large $n$.
For $c>1$, we determine the leading term of the minimum
synthesis time as $(q+1)n/2$ and give lower and upper bounds
on the terms of order $\sqrt n$ using normal quantiles.
For $c=1$, we combine differential VT codes
\cite{NguyenDVT2024} with binary constant-weight codes
to reduce the general upper bound $qn$ by an amount
proportional to $n$.
The case $\log_q(q-1)\leq c<1$ requires further study.
\end{enumerate}

The remainder of this paper is organized as follows.
In Section \ref{sec:pre}, we introduce the notation
and synthesis model and establish the finite-length bounds.
The probabilistic tools and main results are presented
in Section \ref{sec:main}, followed by their proofs
in Section \ref{sec:proof}.
We conclude in Section \ref{sec:con} with a summary
of the results and a discussion of future work.

\section{Preliminaries}
\label{sec:pre}
This section gives the notations that will be used in the paper.
\subsection{Notations}
Let $q\geq 2$ be an integer, and let
$\Sigma_q=\{0,1,\ldots,q-1\}$ denote the \emph{$q$-ary alphabet}. Unless otherwise specified, in this paper, we set the (shifted) modulo operator  $a\bmod q\in \Sigma_q$. Let $[i, j]$ denote the set $\{i, i+1, \ldots, j\}$ if $i \le j$, and the empty set otherwise. Specifically, when $i = 1$, we abbreviate $[1, j]$ as $[j]$.
The set of all length-$n$ sequences over $\Sigma_q$ is denoted
by $\Sigma_q^n$. A \emph{sequence} in $\Sigma_q^n$ is written as
$\bm{x}=x_1x_2\cdots x_n$, where $x_i\in\Sigma_q$ for every
$i\in[n]$. The length of $\bm{x}$ is denoted by
$\lvert \bm{x}\rvert$.  In addition, for any sequence $\bm{x}\in \Sigma_q^*$, we define a \emph{run} in it as a maximal constant substring, and $r(\bm{x})$ denotes the number of runs of $\bm{x}$.

For any sequences $\bm{x},\S\in \Sigma_q^*$, we say $\S$ is a \emph{supersequence} of $\bm{x}$ if there exist indices
$1\leq i_1<i_2<\cdots<i_{\lvert\bm{x}\rvert}\leq \lvert\S\rvert$ such that
$x_j=s_{i_j}$ for every $j\in[\lvert\bm{x}\rvert]$. Equivalently,
$\bm{x}$ is a \emph{subsequence} of $\S$, which is denoted by
$\bm{x}\preceq\S$. A sequence $\bm{x}$ can be synthesized by
$\S$ if and only if
$\bm{x}$ is a subsequence of $\S$. For a reference sequence $\S$,
let
$\Sub_n(\S)
=\{x\in\Sigma_q^n:x\preceq\S\}$
denote the set of all length-$n$ sequences that can be
synthesized by $\S$.

A $q$-ary code of length $n$ is a set
$\C\subseteq\Sigma_q^n$. Each sequence in
$\C$ is called a codeword, and
$\lvert\C\rvert$ is referred to as the code size.
For $\C\neq\varnothing$, its \emph{redundancy} is defined by
$red(\C)\triangleq n\log_2q-\log_2\lvert\C\rvert$. The code
$\C$ can be synthesized by $\S$ if every
codeword in $\C$ can be synthesized by
$\S$, or equivalently,
$\C\subseteq\Sub_n(\S)$. For $t\in[0,T]$, let $\S_{1:t}$ denote the prefix of $\S$ of length $t$. For a nonempty code $\C\subseteq\Sub_n(\S)$, its \emph{synthesis time} with respect to $\S$ is $t_{\S}(\C)\triangleq\min\{t\in[0,T]:\C\subseteq\Sub_n(\S_{1:t})\}$. For a single sequence, we abbreviate $t_{\S}(\{\bm{x}\})$ as $t_{\S}(\bm{x})$; it is $+\infty$ if $\bm{x}$ is not a subsequence of $\S$. 

For a sequence $\bm{x}\in\Sigma_q^n$ and integer $t\ge 0$, let $\D_t(\bm{x})$ denote the set
of all distinct length-$(n-t)$ sequences obtained by deleting
exactly $t$ symbols from $\bm{x}$. In particular, $\D_0(\bm{x})=\{\bm{x}\}$, and $\D_t(\bm{x})=\varnothing$ if $t>n$. Moreover, for a set $\A$, we denote $\D_t(\A)=\bigcup_{\bm{x}\in\A}\D_t(\bm{x})$. A code
$\C\subseteq\Sigma_q^n$ is called a \emph{single-deletion correcting code} if
$\D_1(\bm{x})\cap \D_1(\bm{x}')=\varnothing$ for every pair of distinct
codewords $\bm{x},\bm{x}'\in\C$.

\subsection{Problem Formulation}
In practical DNA synthesis, the alphabet is quaternary.
For convenience, however, we work with a general $q$-ary alphabet, where $q\geq 2$. Our examples will focus on the quaternary case, in which we use the alphabet
$\{0,1,2,3\}$ in place of
$\{\mathtt{A},\mathtt{T},\mathtt{C},\mathtt{G}\}$. For a code
$\C\subseteq\Sigma_q^n$ and a reference sequence $\S\in\Sigma_q^*$,
the machine synthesizes the strands in $\C$ in parallel by
processing $\S$ from left to right. In the $i$th synthesis cycle,
the symbol $s_i$ may be appended to any selected subset of
the strands being synthesized. The synthesis process continues until every codeword in $\C$
has been fully synthesized. Since one symbol of
$\S$ is processed in each cycle, the synthesis time
is the length of the reference prefix used to complete the code.
Before stating our problems, we need the following two definitions.
\begin{definition}
For an integer $1\leq M\leq q^n$, the minimum synthesis time among all
$q$-ary length-$n$ codes of size $M$ is defined as
\begin{align*}
\tau_q(n,M)
&\triangleq
\min\Bigl\{
\lvert\S\rvert:
\exists\,\C\subseteq\Sub_n(\S)
\text{ satisfies }\lvert\C\rvert=M
\Bigr\},
\end{align*} 
\end{definition}

\begin{definition}
 For an integer $1\leq M\leq q^n$, the minimum synthesis time among all $q$-ary
single-deletion correcting codes of length $n$ and size $M$
is defined as
\begin{align*}
\tau_q^{\del}(n,M)
&\triangleq
\min\Bigl\{
\lvert\S\rvert:
\exists\,\C\subseteq\Sub_n(\S)
\text{ with }\lvert\C\rvert=M,\text{ and}\\
&~~~~~~~~~~~~~~~~~~~~~~~~~~~~
\C~\text{can correct one deletion}
\Bigr\}.
\end{align*} 
 If no such code exists, we set $\tau_q^{\del}(n,M)=+\infty$.
\end{definition}

In this paper, we focus on studying the sizes of $\tau_q(n,M)$ and $\tau_q^{\del}(n,M)$ under different $M$.

\subsection{Tools and Basic Results}\label{subsec:tool}

\subsubsection{Alternating Sequence}
Let $\A_q=a_1a_2a_3\cdots$ denote the infinite
\emph{alternating sequence} over $\Sigma_q$, where
$a_i=(i-1)\bmod q$ for every integer $i\geq1$.
Then, for any integer $T\geq0$, let
$\A_{q,T}=a_1a_2\cdots a_T\in\Sigma_q^T$ denote the prefix
of $\A_q$ of length $T$. Furthermore, for integers $n,T\geq0$, we define
$L_q(n,T)\triangleq\lvert\Sub_n(\A_{q,T})\rvert$, i.e., $L_q(n,T)$ is the number of length-$n$ sequences that
can be synthesized by $\A_{q,T}$. In particular, $L_q(n,T)=0$
when $T<n$.

Then, we review two conclusions about the alternating sequence.

\begin{lemma}\cite[Lemma 2]{Lenz2020}
  For any integers $T\ge 0$ and $0\le n\le T$, and all $\S\in\Sigma_q^T$, it holds that 
  \begin{align*} \lvert\Sub_n(\S)\rvert=\lvert\D_{T-n}(\S)\rvert.  
  \end{align*}
\end{lemma}

\begin{lemma}\cite[Theorem 2.4]{HirschbergRegnier2000}
For integers $q\ge 1$, $T\geq 0$ and $0\leq t\leq T$, and every
$\S\in\Sigma_q^T$, we have
\begin{align*}
   \lvert\D_t(\S)\rvert\leq \lvert\D_t(\A_{q,T})\rvert. 
\end{align*}   
\end{lemma}

From these two lemmas, we can immediately obtain the following conclusion.

\begin{corollary}\label{cor:max-ref}
   For any integers $T\ge 0$ and $0\le n\le T$, and all $\S\in\Sigma_q^T$, it holds that 
   \begin{align*} \lvert\Sub_n(\S)\rvert\le L_q(n,T).  
  \end{align*}
\end{corollary}

\subsubsection{Differential and Interval Vectors}
In this paper, we need two forms of differential sequence, which are defined as follows.
\begin{definition}
  For any sequence $\bm{x}=x_1x_2\cdots x_n\in\Sigma_q^n$, the \emph{differential vector} of $\bm{x}$ is defined as $\Diff(\bm{x})=y_1y_2\cdots y_n$, where
  $y_i=(x_i-x_{i+1})\bmod q$ for $1\leq i<n$ and
  $y_n=x_n$.  
\end{definition}

\begin{definition}
   For any sequence $\bm{x}=x_1x_2\cdots x_n\in\Sigma_q^n$, the \emph{interval vector} of $\bm{x}$ is defined as $\Int(\bm{x})=w_1w_2\cdots w_n$, where $w_i=(x_i-x_{i-1})\bmod q\in [q]$ for $1\leq i\leq n$, and $x_0=q-1$. It should be noted that in this definition, the residue $0$ is represented by $q$.
\end{definition}
We note that the value $w_i$ is the number of
synthesis cycles required to reach $x_i$ after $x_{i-1}$ if the reference sequence is alternating sequence. Thus, the following lemma appears natural.
\begin{lemma}\cite[Lemma 4]{Lenz2020}\label{lem:ref-time}
    For any $\bm{x}=x_1x_2\cdots x_n\in\Sigma_q^n$, we have
    $\bm{x}\in\Sub_n(\A_{q,T})$ if and only if
    $\sum_{i=1}^{n}\Int(\bm{x})_i\leq T$. In this case,
    \begin{align*}
        t_{\A_{q,T}}(\bm{x})=\sum_{i=1}^{n}\Int(\bm{x})_i.
    \end{align*}
\end{lemma}

\subsubsection{Single-deletion Correcting Codes}
We first review the construction of the $q$-ary differential
VT codes, which can correct one deletion.

\begin{lemma}\cite[Theorem 3]{NguyenDVT2024}\label{lem:diffVT}
   For any $q\geq 2$, and $a\in[0,qn-1]$, define
\begin{align*}
\mathrm{DiffVT}_a(n;q)
\triangleq
\left\{
\bm{x}\in\Sigma_q^n:
\sum_{i=1}^{n}i\Diff(\bm{x})_i\equiv a\pmod{qn}
\right\}.
\end{align*}
 The code $\mathrm{DiffVT}_a(n;q)$ can correct a single
deletion or a single insertion in linear time. 
\end{lemma}

Next, we review several conclusions about single-deletion correcting codes.

\begin{lemma}\cite{Levenshtein1965BinaryCC}\label{lem:del-run}
    For any $\bm{x}\in \Sigma_q^*$, it holds that $\lvert\D_1(\bm{x})\rvert=r(\bm{x})$.
\end{lemma}

\begin{lemma}\cite[Theorem 5.1]{KulkarniKiyavash2013}\label{lem:del-up}
   Let $n\geq 2$ and $\A\subseteq\Sigma_q^n$. Every code
   $\C\subseteq\A$ that corrects one deletion satisfies
 \begin{align*}
\lvert\C\rvert
\leq
\sum_{\bm{y}\in\D_1(\A)}
\frac{1}{\lvert\D_1(\bm{y})\rvert}.
\end{align*} 
\end{lemma}

By Lemmas \ref{lem:del-run} and \ref{lem:del-up}, we can obtain a corollary.

\begin{corollary}\label{cor:del-up}
Let $n\geq2$ and let $\S\in \Sigma_q^*$ be a reference sequence. If
$\C\subseteq\Sub_n(\S)$ can correct one deletion, then
\begin{align*}
\sum_{\bm{x}\in\C}r(\bm{x})
&\leq
\left\lvert\Sub_{n-1}(\S)\right\rvert,
\end{align*}
and
\begin{align*}
\lvert\C\rvert
&\leq
\sum_{\bm{y}\in\Sub_{n-1}(\S)}
\frac{1}{r(\bm{y})}.
\end{align*}
\end{corollary}

\begin{proof}
For every $\bm{x}\in\C$, the inclusion
$\C\subseteq\Sub_n(\S)$ implies
$\D_1(\bm{x})\subseteq\Sub_{n-1}(\S)$.
Next, the single-deletion correcting property of $\C$ gives
$\D_1(\bm{x})\cap\D_1(\bm{x}')=\varnothing$ for all distinct
$\bm{x},\bm{x}'\in\C$. Combining with Lemma \ref{lem:del-run}, we can obtain that
\begin{align*}
\sum_{\bm{x}\in\C}r(\bm{x})
&=
\sum_{\bm{x}\in\C}\left\lvert\D_1(\bm{x})\right\rvert
=
\left\lvert\bigcup_{\bm{x}\in\C}\D_1(\bm{x})\right\rvert
\leq
\left\lvert\Sub_{n-1}(\S)\right\rvert.
\end{align*}

For the second inequality, set
$\A=\Sub_n(\S)$, then
$\D_1(\A)\subseteq\Sub_{n-1}(\S)$.
Moreover, Lemma \ref{lem:del-run} gives
$\lvert\D_1(\bm{y})\rvert=r(\bm{y})$ for every
$\bm{y}\in\D_1(\A)$. The Lemma \ref{lem:del-up} then implies
\begin{align*}
\lvert\C\rvert
&\leq
\sum_{\bm{y}\in\D_1(\A)}
\frac{1}{\left\lvert\D_1(\bm{y})\right\rvert}
=
\sum_{\bm{y}\in\D_1(\A)}
\frac{1}{r(\bm{y})}
\leq
\sum_{\bm{y}\in\Sub_{n-1}(\S)}
\frac{1}{r(\bm{y})}.
\end{align*}
\end{proof}

\subsection{The Value of $\tau_q(n,M)$ and
Bounds on $\tau_q^{\del}(n,M)$}

In this subsection, we will provide the implicit value of $\tau_q(n,M)$ and upper and lower bounds of $\tau_q^{\del}(n,M)$ by using the conclusions in Section \ref{subsec:tool}. We first define the inverse of $L_q(n,T)$.

\begin{definition}\label{def:inv}
For integers $n\geq1$ and $K\geq1$, we define
\begin{align*}
T(L_q(n,T),K)
&\triangleq
\min\left\{
T\geq n:
L_q(n,T)\geq K
\right\}.
\end{align*}
In other words, $T(L_q(n,T),K)$ is the minimum length
$t$ such that the alternating reference $\A_{q,t}$ can
synthesize at least $K$ distinct sequences of length $n$.
If $K>q^n$, we set $T(L_q(n,T),K)=+\infty$.
\end{definition}

For every integer $0\leq N\leq q^n$, define
\begin{align*}
W_{q,n}(N)
&\triangleq
\max_{\substack{
\B\subseteq\Sigma_q^n\\
\lvert\B\rvert\leq N}}
\sum_{\bm{y}\in\B}
\frac{1}{r(\bm{y})}.
\end{align*}
We note that, $W_{q,n}(0)=0$, and $W_{q,n}(N)$ is
nondecreasing in $N$. Moreover, an integer $M$ is called \emph{feasible} if there exists a
$q$-ary single-deletion correcting code of length $n$ that has size
$M$. Then, we give the value of $\tau_q(n,M)$ and  bounds of $\tau_q^{\del}(n,M)$ by using $W_{q,n}(N)$ and Lemma \ref{lem:diffVT}. 

\begin{theorem}\label{the:bound}
Let $q\geq2$ and $n\geq1$ be integers.
For every integer $1\leq M\leq q^n$, the following
results hold.
\begin{enumerate}
\item[(i)]
$\tau_q(n,M)=
T(L_q(n,T),M)$.

\item[(ii)]
If $n\geq2$ and $M$ is feasible, then
\begin{align*}
\min\left\{
T\geq n:
W_{q,n-1}\bigl(L_q(n-1,T)\bigr)\geq M
\right\}
&\leq
\tau_q^{\del}(n,M)\leq
\min\left\{
qn,T(L_q(n,T),qnM)
\right\}.
\end{align*}
Moreover, for any $\S\in\Sigma_q^T$ and single-deletion
correcting code $\C\subseteq\Sub_n(\S)$ with
$\lvert\C\rvert=M$, we have
$M\leq W_{q,n-1}(L_q(n-1,T))$.
\end{enumerate}
\end{theorem}

\begin{proof}
For part (i), let $T\geq n$. For any
$\S\in\Sigma_q^T$ and code $\C\subseteq\Sub_n(\S)$
with $\lvert\C\rvert=M$, we obtain from
Corollary \ref{cor:max-ref} that
\begin{align*}
M
=
\lvert\C\rvert
&\leq
\lvert\Sub_n(\S)\rvert
\leq
L_q(n,T).
\end{align*}
By Definition \ref{def:inv}, we have
$T\geq T(L_q(n,T),M)$, so
$\tau_q(n,M)\geq T(L_q(n,T),M)$.
For the reverse inequality, take
$t_1\triangleq T(L_q(n,T),M)$. From the same definition, we know that
\begin{align*}
\left\lvert\Sub_n(\A_{q,t_1})\right\rvert
&=
L_q(n,t_1)
\geq M.
\end{align*}
Therefore, we can choose a code
$\C\subseteq\Sub_n(\A_{q,t_1})$ with $\lvert\C\rvert=M$,
which implies
$\tau_q(n,M)\leq t_1=T(L_q(n,T),M)$.
This proves part (i).

We next prove the upper bound for feasible $M$.
Let $\C\subseteq\Sigma_q^n$ be a single-deletion
correcting code with $\lvert\C\rvert=M$.
It is clear that
$\C\subseteq\Sigma_q^n=\Sub_n(\A_{q,qn})$. 
By the definition of $\tau_q^{\del}(n,M)$, we obtain
$\tau_q^{\del}(n,M)\leq qn$. On the other hand, if $T(L_q(n,T),qnM)=+\infty$, this already gives the
required upper bound. Otherwise, take
$t_2=T(L_q(n,T),qnM)$.
By Definition \ref{def:inv}, we know that
$L_q(n,t_2)\geq qnM$.
Recall that the $q$-ary differential VT codes
$\mathrm{DiffVT}_a(n;q)$ form a partition of $\Sigma_q^n$,
where $a\in[0,qn-1]$.
By the pigeonhole principle, there exists
$a\in[0,qn-1]$ such that
\begin{align*}
\left\lvert
\Sub_n(\A_{q,t_2})\cap\mathrm{DiffVT}_a(n;q)
\right\rvert
&\geq
\frac{L_q(n,t_2)}{qn}
\geq M.
\end{align*}
Choose a set
$\C'\subseteq\Sub_n(\A_{q,t_2})\cap\mathrm{DiffVT}_a(n;q)$
with $\lvert\C'\rvert=M$.
From Lemma \ref{lem:diffVT}, we know that $\C'$ is
a single-deletion correcting code.
Moreover, $\C'\subseteq\Sub_n(\A_{q,t_2})$, so
\begin{align*}
\tau_q^{\del}(n,M)
&\leq t_2
=T(L_q(n,T),qnM).
\end{align*}
Together with $\tau_q^{\del}(n,M)\leq qn$, this proves
the upper bound for all $n\geq1$.

For the lower bound, assume $n\geq2$.
Consider any $\S\in\Sigma_q^T$ and single-deletion
correcting code $\C\subseteq\Sub_n(\S)$ with
$\lvert\C\rvert=M$.
From Corollaries \ref{cor:max-ref} and \ref{cor:del-up},
together with the definition and monotonicity of
$W_{q,n-1}(N)$, we have
\begin{align*}
M
=
\lvert\C\rvert
&\leq
\sum_{\bm{y}\in\Sub_{n-1}(\S)}\frac{1}{r(\bm{y})}
\leq
W_{q,n-1}\bigl(\lvert\Sub_{n-1}(\S)\rvert\bigr)\leq
W_{q,n-1}\bigl(L_q(n-1,T)\bigr).
\end{align*}
By the definition of $\tau_q^{\del}(n,M)$, we obtain
the lower bound in part (ii).
\end{proof}

We note that the value of $\tau_q(n,M)$ and the bounds on $\tau_q^{\del}(n,M)$ all depend on
$L_q(n,T)$.
Therefore, to obtain explicit asymptotic expressions for
the synthesis time, we next estimate $L_q(n,T)$ and use
these estimates to analyze the results in
Theorem \ref{the:bound}.

\section{Main Results}
\label{sec:main}

In this section, we introduce the probabilistic tools
and present the main results of this paper.
For the analysis below, we denote the derivative of a
differentiable function $f$ at $x$ by $f'(x)$. In addition, for $g_n>0$, the notation $f_n=O_a(g_n)$ means that
$\lvert f_n\rvert\leq C_ag_n$ for all sufficiently large $n$,
where $C_a>0$ may depend on the fixed parameter $a$
but not on $n$.
Similarly, $f_n=o_a(g_n)$ means that $f_n/g_n\to0$
for each fixed $a$, with a rate of convergence that may
depend on $a$.
Multiple subscripts indicate dependence on all the
listed parameters.

\subsection{Probabilistic Tools for Estimating $L_q(n,T)$}\label{subsec:prob}

The value of $\tau_q(n,M)$ and the upper and lower bounds
on $\tau_q^{\del}(n,M)$ in Theorem \ref{the:bound} depend on
$L_q(n,T)$. We first express $L_q(n,T)/q^n$ as a probability
involving a sum of i.i.d.\ uniform random variables.
To estimate this probability, we then introduce the
exponentially tilted distribution, the log-moment-generating
function, and the Cram\'er rate function.

\subsubsection{Uniform Distribution}

Let $\bm{X}$ be a uniform random variable over $\Sigma_q^n$,
and define
$\bm{U}\triangleq\Int(\bm{X})=(U_1,\ldots,U_n)$.
Since the map $\Int:\Sigma_q^n\to[q]^n$ is a bijection,
$U_1,\ldots,U_n$ are i.i.d.\ uniform random variables over
$[q]$. The probability mass function of each $U_i$ is given by
$\Prob\{U_i=j\}=1/q$ for every $i\in[n]$ and $j\in[q]$.

Let $Z_n\triangleq\sum_{i=1}^n U_i$. By Lemma
\ref{lem:ref-time}, we have
$\bm{x}\in\Sub_n(\A_{q,T})$ if and only if
$\sum_{i=1}^n\Int(\bm{x})_i\leq T$. Recall that
$L_q(n,T)=\lvert\Sub_n(\A_{q,T})\rvert$. Therefore,
\begin{align*}
\frac{L_q(n,T)}{q^n}
&=
\frac{1}{q^n}
\left\lvert
\left\{
\bm{x}\in\Sigma_q^n:
\sum_{i=1}^n\Int(\bm{x})_i\leq T
\right\}
\right\rvert
=
\Prob\{Z_n\leq T\}.
\end{align*}
Furthermore, the problem of estimating
$L_q(n,T)$ can be reduced to estimating $\Prob\{Z_n\leq T\}$. 

We know that the mean and variance of the uniform distribution
on $[q]$ are $(q+1)/2$ and $(q^2-1)/12$, respectively.
Accordingly, we define
\begin{align*}
\mu_q
&\triangleq
\E[U_i]
=
\frac{q+1}{2},
\quad
\sigma_q^2
\triangleq
\Var(U_i)
=
\frac{q^2-1}{12}.
\end{align*}
Since $U_1,\ldots,U_n$ are i.i.d.\ with mean $\mu_q$ and
variance $\sigma_q^2$, we have
\begin{align*}
\E[Z_n]
&=
n\mu_q,
\quad
\Var(Z_n)
=
n\sigma_q^2.
\end{align*}
Thus, the standard deviation of $Z_n$ is
$\sigma_q\sqrt n$. 

\subsubsection{Exponentially Tilted Distribution}

We now consider $\Prob\{Z_n\leq T\}$ when $T$ lies below the
mean $n\mu_q$. Since the smallest possible value of $Z_n/n$
is $1$, we consider $T/n\in(1,\mu_q)$.

To estimate $\Prob\{Z_n\leq T\}$, we introduce auxiliary
random variables by reweighting the probabilities
$\Prob\{U_i=j\}$. Let $\mathbb R$ denote the set of real
numbers. For $\theta\in\mathbb R$, the
\emph{exponentially tilted distribution} on $[q]$ has
probability mass function
\begin{align*}
p_\theta(j)
&\triangleq
\frac{e^{\theta j}\Prob\{U_i=j\}}
{\E[e^{\theta U_i}]}
=
\frac{e^{\theta j}}
{\sum_{a=1}^q e^{\theta a}},
\end{align*}
where $j\in[q]$. Let $U_1^\theta,\ldots,U_n^\theta$ be
i.i.d.\ random variables satisfying
$\Prob\{U_i^\theta=j\}=p_\theta(j)$ for every $i\in[n]$
and $j\in[q]$. We note that, when $\theta=0$, $p_0(j)=1/q$ for every
$j\in[q]$, so $U_i^0$ has the same distribution as $U_i$.

\subsubsection{Log-Moment-Generating Function}

We next introduce a function whose derivatives give the
mean and variance of $U_i^\theta$. This function will also
be used to define the rate function.
For $\theta\in\mathbb R$, define the
\emph{log-moment-generating function} of $U_i$ by
\begin{align*}
\psi_q(\theta)
&\triangleq
\ln\E[e^{\theta U_i}]
=
\ln\left(
\frac{1}{q}\sum_{j=1}^q e^{\theta j}
\right).
\end{align*}
From the definition of $\psi_q(\theta)$, we have
$\E[e^{\theta U_i}]=e^{\psi_q(\theta)}$. Since
$Z_n=\sum_{i=1}^n U_i$ and $U_1,\ldots,U_n$ are independent,
we obtain
\begin{align*}
\E[e^{\theta Z_n}]
&=
\prod_{i=1}^n\E[e^{\theta U_i}]
=
e^{n\psi_q(\theta)}.
\end{align*}
Moreover, from the definition of $p_\theta(j)$,
we have
\begin{align*}
\psi_q'(\theta)
&=
\frac{\sum_{j=1}^q j e^{\theta j}}
{\sum_{j=1}^q e^{\theta j}}
=
\sum_{j=1}^q j p_\theta(j)
=
\E[U_i^\theta],
\\
\psi_q''(\theta)
&=
\frac{\sum_{j=1}^q j^2 e^{\theta j}}
{\sum_{j=1}^q e^{\theta j}}
-
\left(
\frac{\sum_{j=1}^q j e^{\theta j}}
{\sum_{j=1}^q e^{\theta j}}
\right)^2
\\
&=
\E\bigl[(U_i^\theta)^2\bigr]
-
\bigl(\E[U_i^\theta]\bigr)^2
=
\Var(U_i^\theta)>0.
\end{align*}
Since $q\geq2$ and $p_\theta(j)>0$ for every $j\in[q]$,
$U_i^\theta$ is not constant. The last inequality therefore
follows from the definition of variance.

To estimate $\Prob\{Z_n\leq T\}$, we use probabilities
involving the auxiliary sum $\sum_{i=1}^n U_i^\theta$.
We seek a parameter $\theta$ such that
$\E[\sum_{i=1}^n U_i^\theta]=T$, allowing us to estimate probabilities near its mean
and relate them to the original probability.
Since
$\E[\sum_{i=1}^n U_i^\theta]=n\E[U_i^\theta]$,
we require $\E[U_i^\theta]=T/n$.

The following lemma gives some properties of $\psi_q$
and the corresponding tilting parameter.

\begin{lemma}\label{lem:tilt}
Let $q\geq2$ be fixed and $\mathbb R$ denote the field of real numbers. The following results hold.
\begin{enumerate}
\item[(i)]
The function $\psi_q'$ is continuous and strictly
increasing on $\mathbb R$, with
$\lim_{\theta\to-\infty}\psi_q'(\theta)=1$.
At $\theta=0$, we have
\begin{align*}
\psi_q(0)
&=0,
\quad
\psi_q'(0)=\mu_q,
\quad
\psi_q''(0)=\sigma_q^2,
\quad
\psi_q'''(0)=0.
\end{align*}

\item[(ii)]
For every $\gamma\in(1,\mu_q)$, there exists a unique
\emph{tilting parameter} $\theta_q(\gamma)<0$ satisfying
\begin{align*}
\psi_q'\bigl(\theta_q(\gamma)\bigr)
&=
\E\bigl[U_i^{\theta_q(\gamma)}\bigr]
=
\gamma.
\end{align*}
Moreover, $\theta_q(\gamma)\to0$ as
$\gamma\rightarrow\mu_q$.
\end{enumerate}
\end{lemma}

\begin{proof}
For part (i), the continuity of $\psi_q'$ follows
from its finite-sum expression, whose denominator
is positive.
Since $\psi_q''(\theta)>0$, the function $\psi_q'$
is strictly increasing.
The values of $\psi_q(0)$, $\psi_q'(0)$, and
$\psi_q''(0)$ follow from the definition of $\psi_q$
and the mean and variance of $U_i$.
Moreover, $U_i-\mu_q$ and $\mu_q-U_i$ have the same
distribution, so
$\E[e^{\theta(U_i-\mu_q)}]
=\E[e^{-\theta(U_i-\mu_q)}]$. Consequently,
\begin{align*}
\psi_q(\theta)-\mu_q\theta
&=
\ln\E[e^{\theta(U_i-\mu_q)}]
=
\ln\E[e^{-\theta(U_i-\mu_q)}]
=
\psi_q(-\theta)+\mu_q\theta.
\end{align*}
Thus, $\psi_q(\theta)-\mu_q\theta$ is an even function,
which implies $\psi_q'''(0)=0$.
As $\theta\to-\infty$, we have
$p_\theta(1)
=1/(1+\sum_{j=2}^q e^{\theta(j-1)})\to1$.
Then,
\begin{align*}
0\leq\psi_q'(\theta)-1
&=
\sum_{j=2}^q(j-1)p_\theta(j)
\leq
(q-1)\bigl(1-p_\theta(1)\bigr)
\longrightarrow0.
\end{align*}

For part (ii), by the continuity and strict
monotonicity of $\psi_q'$, together with
$\psi_q'(0)=\mu_q$ and
$\lim_{\theta\to-\infty}\psi_q'(\theta)=1$,
we know that $\psi_q'(\theta)=\gamma$ has a unique
negative solution for every $\gamma\in(1,\mu_q)$.
We denote this solution by $\theta_q(\gamma)$.
The same continuity and monotonicity, together with
$\psi_q'(0)=\mu_q$, imply that
$\theta_q(\gamma)\to0$ as $\gamma\rightarrow\mu_q$.
\end{proof}

For the probability $\Prob\{Z_n\leq T\}$, the prescribed
mean is $T/n$, so we take $\gamma=T/n$.

\subsubsection{Cram\'er Rate Function}

To study the rate of decay of $\Prob\{Z_n\leq T\}$,
we introduce the \emph{Cram\'er rate function}.
For $\gamma\in(1,\mu_q)$, we define the
\emph{Cram\'er rate function} by
\begin{align*}
I_q(\gamma)
&\triangleq
\sup_{\theta\in\mathbb R}
\left\{
\theta\gamma-\psi_q(\theta)
\right\}.
\end{align*}

The following lemma gives some properties of the
Cram\'er rate function $I_q$.

\begin{lemma}\label{lem:rate}
Let $q\geq2$ be fixed. The following results hold.
\begin{enumerate}
\item[(i)]
For every $\gamma\in(1,\mu_q)$, we have
\begin{align*}
I_q(\gamma)
&=
\gamma\theta_q(\gamma)
-\psi_q\bigl(\theta_q(\gamma)\bigr),
\quad
I_q'(\gamma)=\theta_q(\gamma)<0.
\end{align*}
Moreover, $I_q''(\gamma)=
\theta_q'(\gamma)=
\frac{1}{\psi_q''(\theta_q(\gamma))}>0.$

\item[(ii)]
The function $I_q$ is continuous and strictly decreasing
from $\ln q$ to $0$ on $(1,\mu_q)$.
Consequently, $I_q(\gamma)\to0$ if and only if
$\gamma\to\mu_q$ within this interval.
Furthermore,
$\lim_{\gamma\rightarrow\mu_q}I_q''(\gamma)
=1/\sigma_q^2$.
In particular, $I_q''$ and $\theta_q'$ are bounded
in a left neighborhood of $\mu_q$.
\end{enumerate}
\end{lemma}

\begin{proof}
For part (i), the first and second derivatives of
$\theta\gamma-\psi_q(\theta)$ with respect to $\theta$
are $\gamma-\psi_q'(\theta)$ and
$-\psi_q''(\theta)<0$, respectively.
By Lemma \ref{lem:tilt}(ii), we have
$\gamma-\psi_q'(\theta)=0$ if and only if
$\theta=\theta_q(\gamma)$.
Therefore, the supremum in the definition of
$I_q(\gamma)$ is attained uniquely at $\theta_q(\gamma)$,
which gives the stated expression for $I_q(\gamma)$.
Since $\psi_q''(\theta)>0$, the inverse function
$\theta_q$ is differentiable.
By differentiating
$\psi_q'(\theta_q(\gamma))=\gamma$
with respect to $\gamma$, we obtain
$\psi_q''(\theta_q(\gamma))\theta_q'(\gamma)=1$.
Moreover,
\begin{align*}
I_q'(\gamma)
&=
\theta_q(\gamma)
+\bigl(\gamma-\psi_q'(\theta_q(\gamma))\bigr)
\theta_q'(\gamma)
=
\theta_q(\gamma),
\end{align*}
which gives the stated expression for $I_q''(\gamma)$.

For part (ii), from Lemma \ref{lem:tilt}, we have
$\theta_q(\gamma)\to0$ as $\gamma\rightarrow\mu_q$
and $\psi_q(0)=0$.
Together with the expression for $I_q(\gamma)$
in part (i), we obtain $I_q(\gamma)\to0$.
On the other hand, as $\theta\to-\infty$, we have
\begin{align*}
\psi_q(\theta)
&=
\ln\left(
\frac{e^\theta}{q}
\left(1+\sum_{j=2}^q e^{\theta(j-1)}\right)
\right)
=
\theta-\ln q+
\ln\left(1+\sum_{j=2}^q e^{\theta(j-1)}\right).
\end{align*}
Since $0\leq\ln(1+x)\leq x$ for $x\geq0$, we obtain
\begin{align*}
0\leq
\ln\left(1+\sum_{j=2}^q e^{\theta(j-1)}\right)
&\leq
\sum_{j=2}^q e^{\theta(j-1)}
\leq(q-1)e^\theta.
\end{align*}
Thus, $\psi_q(\theta)=\theta-\ln q+O_q(e^\theta)$.
For $\psi_q'(\theta)$, we have
\begin{align*}
0\leq\psi_q'(\theta)-1
&=
\frac{\sum_{j=2}^q(j-1)e^{\theta(j-1)}}
{1+\sum_{j=2}^q e^{\theta(j-1)}}\leq
e^\theta\sum_{j=2}^q(j-1)
=
\frac{q(q-1)}2e^\theta,
\end{align*}
which gives $\psi_q'(\theta)=1+O_q(e^\theta)$.
By part (i), we obtain $I_q\bigl(\psi_q'(\theta)\bigr)=
\theta\psi_q'(\theta)-\psi_q(\theta)
\longrightarrow\ln q.$
Together with $I_q'(\gamma)<0$, these limits give the
stated decrease of $I_q$ and the equivalence
$I_q(\gamma)\to0$ if and only if $\gamma\to\mu_q$.
Finally, by part (i) and $\psi_q''(0)=\sigma_q^2>0$,
we obtain $\lim_{\gamma\to\mu_q}I_q''(\gamma)=1/\sigma_q^2$.
Thus, $I_q''$ and $\theta_q'$ are bounded in a left
neighborhood of $\mu_q$.
\end{proof}

The notation used in this subsection is summarized
in Table \ref{tab:prob}.

\begin{table*}[t]
\caption{Notation for the probabilistic analysis}
\label{tab:prob}
\centering
\small
\renewcommand{\arraystretch}{1.2}
\setlength{\tabcolsep}{5pt}
\begin{tabular}{
@{}
>{\centering\arraybackslash}p{0.16\textwidth}
>{\raggedright\arraybackslash}p{0.38\textwidth}
>{\raggedright\arraybackslash}p{\dimexpr0.46\textwidth-4\tabcolsep\relax}
@{}
}
\toprule
Notation & Definition & Remarks \\
\midrule

$\bm{X}$
&
A random variable uniformly distributed over $\Sigma_q^n$.
&
$\Prob\{\bm{X}=\bm{x}\}=q^{-n}$ for every
$\bm{x}\in\Sigma_q^n$.
\\

$\bm{U},\,U_i$
&
$\bm{U}=\Int(\bm{X})=(U_1,\ldots,U_n)$.
&
$U_1,\ldots,U_n$ are i.i.d.\ uniform random variables
over $[q]$.
\\

$\theta$
&
A real parameter of the auxiliary distribution.
&
---
\\

$p_\theta(j)$
&
$\displaystyle
\frac{e^{\theta j}}{\sum_{a=1}^q e^{\theta a}}$,
where $j\in[q]$.
&
$p_\theta(j)>0$ and $p_0(j)=1/q$.
When $\theta<0$, $p_\theta(j)$ is strictly decreasing
in $j$.
\\

$U_i^\theta$
&
$\Prob\{U_i^\theta=j\}=p_\theta(j)$.
&
$U_1^\theta,\ldots,U_n^\theta$ are i.i.d.
The variables $U_i^0$ and $U_i$ have the same distribution.
\\

\midrule

$\psi_q(\theta)$
&
$\ln\E[e^{\theta U_i}]$.
&
$\psi_q'(\theta)=\E[U_i^\theta]$ and
$\psi_q''(\theta)=\Var(U_i^\theta)>0$.
\\

$\mu_q$
&
$\displaystyle\E[U_i]=\frac{q+1}{2}$.
&
$\psi_q'(0)=\mu_q$.
\\

$\sigma_q^2$
&
$\displaystyle\Var(U_i)=\frac{q^2-1}{12}$.
&
$\psi_q''(0)=\sigma_q^2>0$.
\\

$Z_n$
&
$\displaystyle\sum_{i=1}^n U_i$.
&
$\E[Z_n]=n\mu_q$ and
$\Var(Z_n)=n\sigma_q^2$.
The standard deviation of $Z_n$ is $\sigma_q\sqrt n$.
\\

$L_q(n,T)$
&
$\lvert\Sub_n(\A_{q,T})\rvert$.
&
$\displaystyle\frac{L_q(n,T)}{q^n}
=\Prob\{Z_n\leq T\}$.
\\

\midrule

$\gamma$
&
A prescribed mean in $(1,\mu_q)$.
&
For $T/n\in(1,\mu_q)$, use $\gamma=T/n$
when estimating $\Prob\{Z_n\leq T\}$.
\\

$\theta_q(\gamma)$
&
The tilting parameter for the exponentially tilted
distribution with mean $\gamma$, defined as the unique
negative solution of $\psi_q'(\theta)=\gamma$.
&
$\E[U_i^{\theta_q(\gamma)}]=\gamma$.
For fixed $q$ and $\gamma$, $\theta_q(\gamma)$
is independent of $n$.
\\

$I_q(\gamma)$
&
$\displaystyle
\sup_{\theta\in\mathbb R}
\{\theta\gamma-\psi_q(\theta)\}$.
&
$I_q'(\gamma)<0$ and $I_q''(\gamma)>0$.
The probability estimate contains the factor
$e^{-nI_q(T/n)}$.
\\

\bottomrule
\end{tabular}
\end{table*}

\subsection{Codes with Linear Redundancy}

For a nonempty code $\C\subseteq\Sigma_q^n$, the
\emph{code rate} is defined as
$R(\C)\triangleq\log_q\lvert\C\rvert/n$.
In this subsection, we consider codes with redundancy
proportional to $n$, corresponding to a fixed limiting
code rate in $(0,1)$.
Let $q\geq3$, $R\in(0,1)$, and $A>0$ be fixed.
We define $M_n\triangleq\lfloor Aq^{Rn}\rfloor$.
For these code sizes, the redundancy satisfies $n\log_2q-\log_2M_n=
(1-R)n\log_2q-\log_2A+o(1).$
The following theorem gives the minimum synthesis time
for unrestricted codes and single-deletion correcting
codes of size $M_n$.

\begin{theorem}\label{the:fix}
For $M_n$ defined above, as $n\to\infty$, we have
\begin{align*}
\tau_q^{\del}(n,M_n)
&=
\gamma_{q,R}n
+\frac{3}{2\lvert\theta_{q,R}\rvert}\ln n
+O_{q,R,A}(1),\\
\tau_q(n,M_n)
&=
\gamma_{q,R}n
+\frac{1}{2\lvert\theta_{q,R}\rvert}\ln n
+O_{q,R,A}(1),
\end{align*}
where $\gamma_{q,R}\in(1,\mu_q)$ is the unique solution of
$I_q(\gamma_{q,R})=(1-R)\ln q$, and
$\theta_{q,R}\triangleq
\theta_q(\gamma_{q,R})=I_q'(\gamma_{q,R})<0$.
Consequently,
\begin{align*}
\tau_q^{\del}(n,M_n)-\tau_q(n,M_n)
&=
\frac{\ln n}{\lvert\theta_{q,R}\rvert}
+O_{q,R,A}(1).
\end{align*}
\end{theorem}
For fixed $q$ and $R$, $\lvert\theta_{q,R}\rvert$
is a positive constant independent of $n$.
Thus, the second terms in both synthesis-time expansions
are of order $\ln n$.

\begin{figure}[t]
\centering
\includegraphics[width=0.45\textwidth]{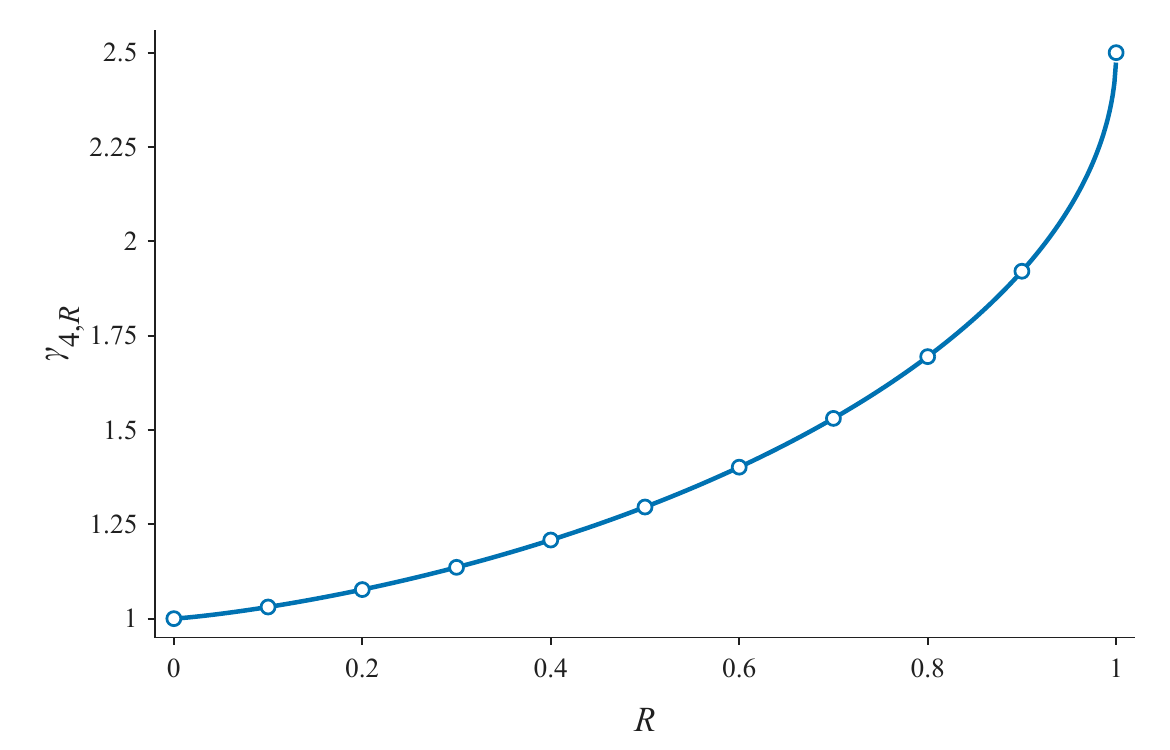}
\caption{This figure shows the relationship between
the code rate $R$ and $\gamma_{4,R}$ in the quaternary case,
where $M_n=\lfloor A4^{Rn}\rfloor$ with fixed $A>0$.
Moreover, $\gamma_{4,R}\in(1,5/2)$ is the unique solution
of $I_4(\gamma_{4,R})=(1-R)\ln4$, and $n\gamma_{4,R}$
is the leading term of the minimum synthesis time
in Theorem \ref{the:fix}.}
\label{fig:rate}
\end{figure}

From Theorem \ref{the:fix}, we see that the
single-deletion correcting constraint adds only a
logarithmic term to the minimum synthesis time. Taking $q=4$ as an example, we illustrate the relationship
between the code rate $R$ and the leading synthesis-time
coefficient $\gamma_{4,R}$ in Fig. \ref{fig:rate}.

\subsection{Codes with Redundancy $\log_2 n+s_n\log_2q+o(1)$}

The preceding subsection considers redundancy proportional
to $n$. We now consider single-deletion correcting codes
whose redundancy exceeds $\log_2 n$ by a quantity that
tends to infinity but is sublinear in $n$.
Let $q\geq3$ be fixed and let $s_n$ satisfy
$s_n\to\infty$ and $s_n=o(n)$.
We define $M_n\triangleq
\left\lfloor\frac{q^n}{nq^{s_n}}\right\rfloor.$
These code sizes correspond to redundancy
$\log_2 n+s_n\log_2q+o(1)$, where the $o(1)$ term comes from
integer rounding.
For these code sizes, we obtain the following bounds on
$\tau_q^{\del}(n,M_n)$.

\begin{figure}[t]
\centering
\includegraphics[width=0.45\textwidth]{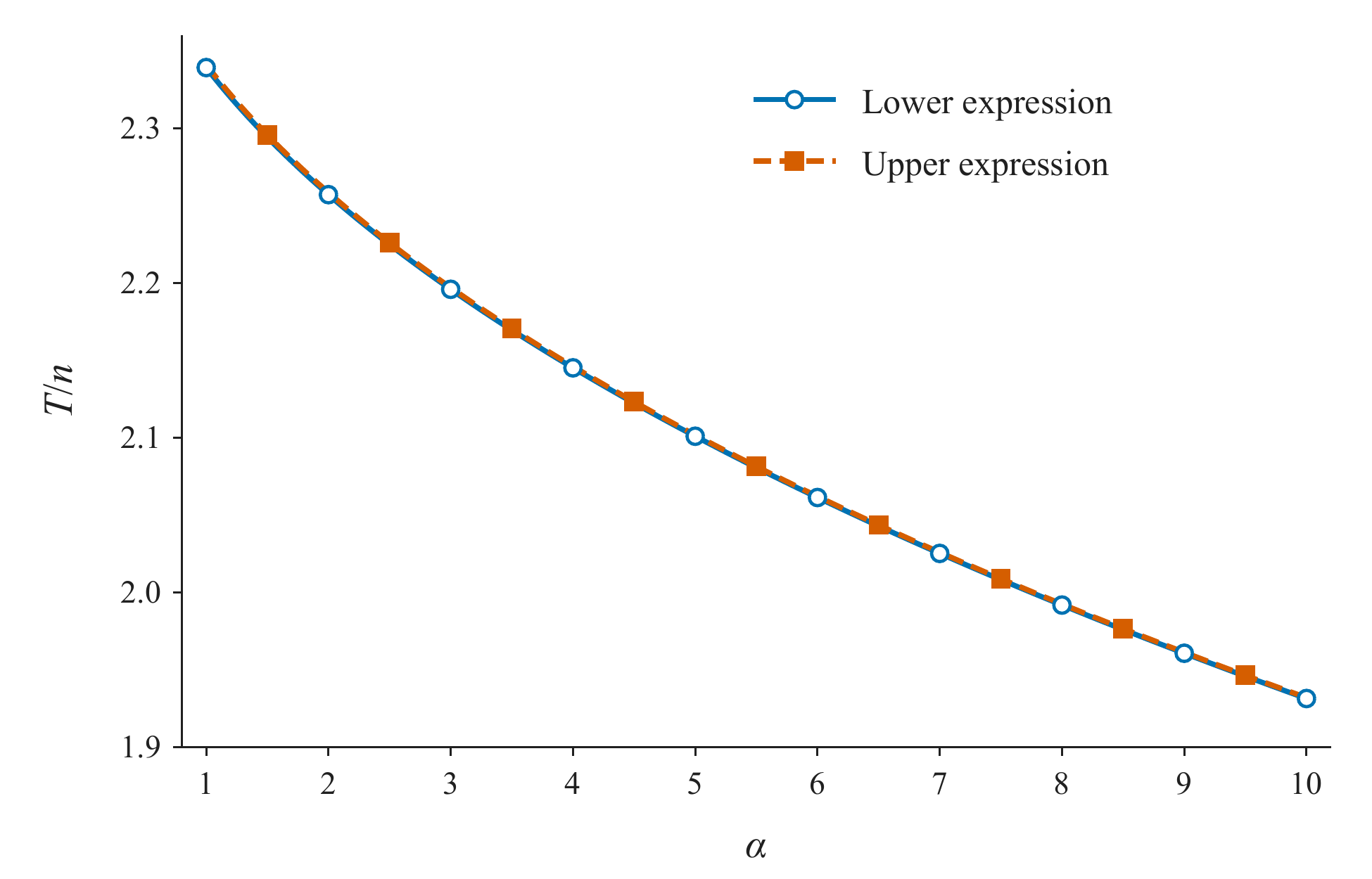}
\caption{In this figure, we compare the normalized expressions
in the lower and upper bounds on the synthesis time given in
Theorem \ref{the:high}, with $q=4$, $n=10^3$, and
$M_n=\lfloor4^n/(n4^{s_n})\rfloor$.
Since the asymptotic bounds require $s_n\to\infty$ and
$s_n=o(n)$, we take $s_n=\alpha\log_2 n$ with fixed
$\alpha\in[1,10]$. The horizontal coordinate is $\alpha$.
The vertical coordinate $T/n$ is
$\gamma_n+[\frac12\ln(4\pi s_n\ln4)+\ln a]/
(n\lvert\theta_n\rvert)$, where
$I_4(\gamma_n)=s_n\ln4/n$ and $\theta_n=\theta_4(\gamma_n)$.
The blue solid curve with open circles uses $a=3$ for the lower bound,
and the orange dashed curve with filled squares uses $a=4$ for the upper bound.
The asymptotic remainder terms are omitted.
The two plotted expressions are close for these code sizes.}
\label{fig:excess}
\end{figure}

\begin{theorem}\label{the:high}
For $M_n$ defined above, as $n\to\infty$, we have
\begin{align*}
\tau_q^{\del}(n,M_n)
&\geq
n\gamma_n
+\frac{\frac12\ln(4\pi s_n\ln q)+\ln(q-1)}
{\lvert\theta_n\rvert}
+o_q\left(\frac1{\lvert\theta_n\rvert}\right),\\
\tau_q^{\del}(n,M_n)
&\leq
n\gamma_n
+\frac{\frac12\ln(4\pi s_n\ln q)+\ln q}
{\lvert\theta_n\rvert}
+o_q\left(\frac1{\lvert\theta_n\rvert}\right),
\end{align*}
where $\gamma_n\in(1,\mu_q)$ is the unique solution of
$I_q(\gamma_n)=s_n\ln q/n$ for all sufficiently large $n$,
and $\theta_n\triangleq\theta_q(\gamma_n)=I_q'(\gamma_n)<0$.
Consequently, the gap between these upper and lower bounds is $\frac{\ln\bigl(q/(q-1)\bigr)}{\lvert\theta_n\rvert}
+o_q\left(\frac1{\lvert\theta_n\rvert}\right).$
\end{theorem}
In the proof of Theorem \ref{the:high}, we can obtain
$1/\lvert\theta_n\rvert=O_q(\sqrt{n/s_n})$.
Hence, for $a\in\{q-1,q\}$, we have
\begin{align*}
\frac{\frac12\ln(4\pi s_n\ln q)+\ln a}
{\lvert\theta_n\rvert}
&=
O_q\left(\sqrt{\frac{n}{s_n}}\ln s_n\right)
=
\sqrt n\,O_q\left(\frac{\ln s_n}{\sqrt{s_n}}\right)
=
o(\sqrt n).
\end{align*}
Thus, the second terms in both the upper and lower bounds
are $o(\sqrt n)$.
Taking $q=4$ and $n=10^3$, Fig. \ref{fig:excess}
shows the relationship between $\alpha$ and the normalized
expressions in the synthesis-time bounds, where
$s_n=\alpha\log_2 n$ and $1\leq\alpha\leq10$.

Under the stronger condition $s_n=o(\sqrt n)$,
we can expand both $\gamma_n$ and $\theta_n$ to obtain
the following bounds expressed directly in terms of $s_n$.

\begin{corollary}\label{cor:high-exp}
Assume the hypotheses of Theorem \ref{the:high}.
If $s_n=o(\sqrt n)$, then, as $n\to\infty$, the lower
and upper bounds in that theorem are given by the
following expression with $a=q-1$ and $a=q$, respectively:
\begin{align*}
\mu_q n
&-\sigma_q\sqrt{2ns_n\ln q}
+\sigma_q\sqrt{\frac{n}{2s_n\ln q}}
\left[\frac12\ln(4\pi s_n\ln q)+\ln a\right]
+o_q\left(\sqrt{\frac{n}{s_n\ln q}}\right).
\end{align*}
\end{corollary}

In particular, by setting $s_n=(r-1)\log_q n$ in
Corollary \ref{cor:high-exp}, we immediately obtain the
following result for redundancy $r\log_2 n+o(1)$,
where $r>1$ is fixed.

\begin{corollary}\label{cor:high-poly}
Let $q\geq3$ and $r>1$ be fixed. We define $M_n\triangleq
\left\lfloor\frac{q^n}{n^r}\right\rfloor.$
Then, as $n\to\infty$, we have
\begin{align*}
\tau_q^{\del}(n,M_n)
&=
\mu_qn-\sigma_q\sqrt{2(r-1)n\ln n}
+\frac{\sigma_q}{2\sqrt{2(r-1)}}
\sqrt{\frac{n}{\ln n}}\ln\ln n
+O_{q,r}\left(\sqrt{\frac{n}{\ln n}}\right).
\end{align*}
\end{corollary}

\subsection{Codes with Redundancy $\log_2 n+c\log_2 q$}

The preceding subsection assumes $s_n\to\infty$ and
therefore does not cover redundancy that differs from
$\log_2 n$ by a constant.
To study this case, let $q\geq3$ and $c\in\mathbb R$
be fixed. We define $M_{n,c}\triangleq
\left\lfloor\frac{q^n}{nq^c}\right\rfloor.$
These code sizes correspond to redundancy
$\log_2 n+c\log_2 q+o(1)$.
Let $\Phi$ denote the cumulative distribution function of
the standard normal distribution, and let $\Phi^{-1}$
denote its inverse. We obtain the following theorem on
feasibility and minimum synthesis time.

\begin{figure}[t]
\centering
\includegraphics[width=0.45\textwidth]{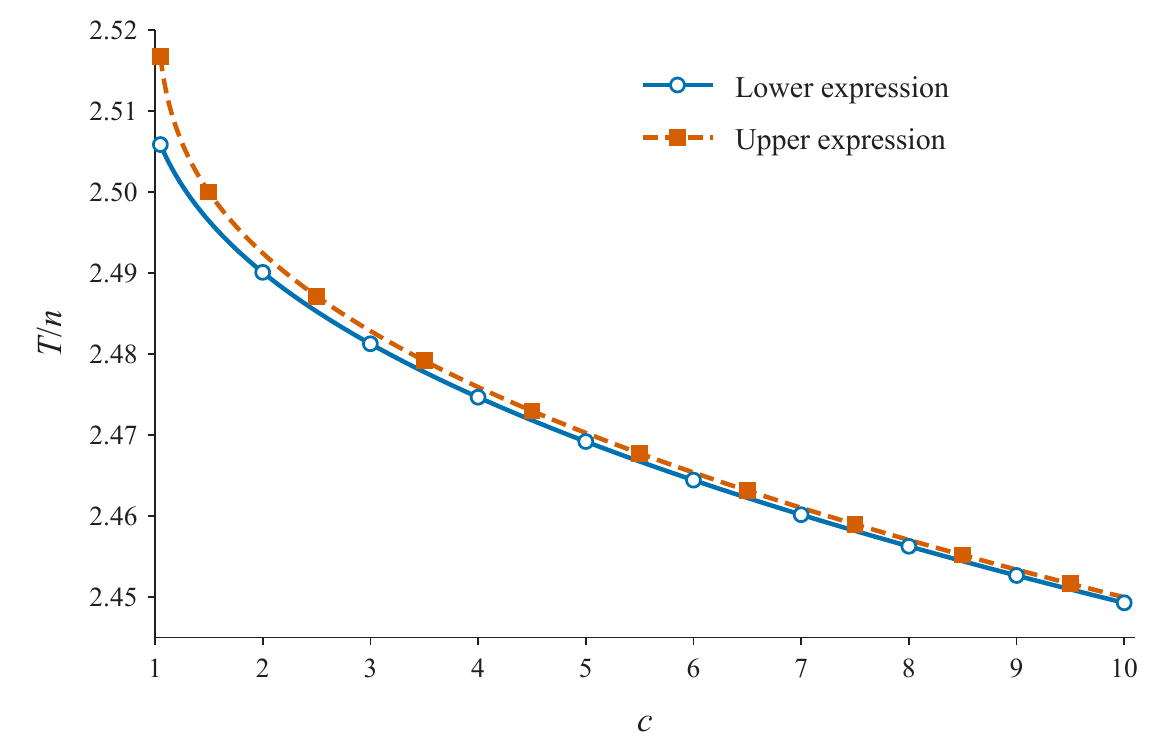}
\caption{In this figure, we compare the normalized expressions
in the lower and upper bounds in Theorem \ref{the:off}, with
$q=4$, $n=10^4$, and $M_{n,c}=\lfloor4^n/(n4^c)\rfloor$
for $1.05\leq c\leq10$.
The blue solid curve with open circles and the orange dashed
curve with filled squares correspond to the lower and upper
bounds, respectively, with the $O(1/n)$ terms omitted.}
\label{fig:offset}
\end{figure}

\begin{theorem}\label{the:off}
For $M_{n,c}$ defined above, the following results hold.
\begin{enumerate}
\item[(i)]
If $c<\log_q(q-1)$, then, for all sufficiently large $n$,
there is no $q$-ary single-deletion correcting code of
length $n$ that has size $M_{n,c}$. Consequently, we have
\begin{align*}
\tau_q^{\del}(n,M_{n,c})
&=+\infty.
\end{align*}

\item[(ii)]
If $c>\log_q(q-1)$, then, as $n\to\infty$, we have
\begin{align*}
\tau_q^{\del}(n,M_{n,c})
&\geq
\mu_qn
+\sigma_q\Phi^{-1}\bigl((q-1)q^{-c}\bigr)\sqrt n
-O_{q,c}(1).
\end{align*}

\item[(iii)]
If $c>1$, then $M_{n,c}$ is feasible for all sufficiently
large $n$, and, as $n\to\infty$, we have
\begin{align*}
\tau_q^{\del}(n,M_{n,c})
&\leq
\mu_qn
+\sigma_q\Phi^{-1}\bigl(q^{1-c}\bigr)\sqrt n
+O_{q,c}(1).
\end{align*}
\end{enumerate}
In particular, for $c>1$, we have
\begin{align*}
\tau_q^{\del}(n,M_{n,c})
&=
\mu_qn+O_{q,c}(\sqrt n),
\quad
\lim_{n\to\infty}
\frac{\tau_q^{\del}(n,M_{n,c})}{n}
=\mu_q.
\end{align*}
\end{theorem}
Taking $q=4$ and $n=10^4$ as an example,
Fig. \ref{fig:offset} shows the relationship between
the redundancy offset $c$ and the normalized expressions
in the synthesis-time bounds.

When $c=1$, the upper bound in Theorem \ref{the:off}(iii)
does not apply. The code size is
$M_{n,1}=\lfloor q^n/(qn)\rfloor$, for which
Theorem \ref{the:bound}(ii) gives
$\tau_q^{\del}(n,M_{n,1})\leq qn$.
The following theorem improves this upper bound.

\begin{theorem}\label{the:c1}
Let $q\geq3$ be fixed. As $n\to\infty$, we have
\begin{align*}
\tau_q^{\del}(n,M_{n,1})
&\leq
(q+1-\rho_q)n
-\frac{\ln n}{2\lvert\theta_q(\rho_q)\rvert}
+O_q(1),
\end{align*}
where $\rho_q\in(1,3/2)$ is the unique solution of
$I_q(\rho_q)=\ln(q/2)$.
\end{theorem}

Since $q-1/2<q+1-\rho_q<q$, this bound reduces the
coefficient of the linear term in the upper bound $qn$.
The first-order value of $\tau_q^{\del}(n,M_{n,1})$
remains undetermined.

\begin{figure}[t]
\centering
\includegraphics[width=0.45\textwidth]{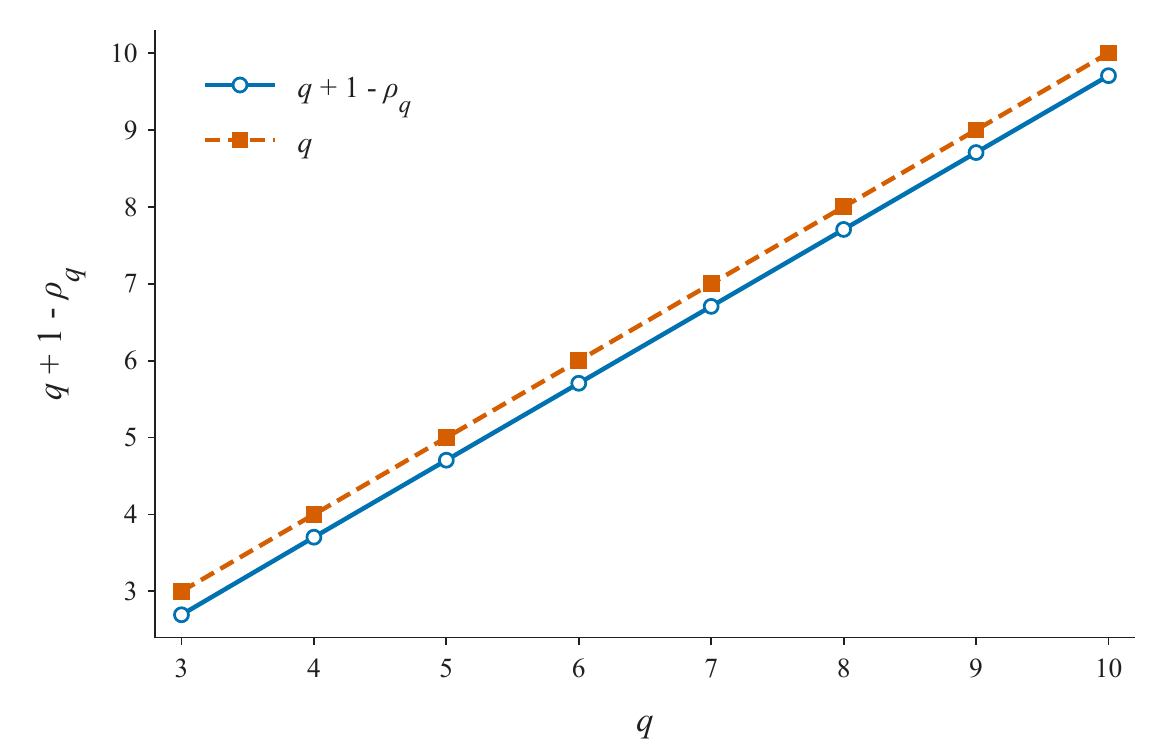}
\caption{This figure shows the relationship between $q$
and $q+1-\rho_q$ for $M_{n,1}=\lfloor q^n/(qn)\rfloor$
and integer values $3\leq q\leq10$.
The solid curve with open circles represents the coefficient
$q+1-\rho_q$ in the upper bound of Theorem \ref{the:c1},
where $I_q(\rho_q)=\ln(q/2)$.
The dashed curve with filled squares represents the
coefficient $q$ in the general upper bound
$\tau_q^{\del}(n,M_{n,1})\leq qn$.}
\label{fig:c1}
\end{figure}

The relationship between $q$ and the upper-bound
coefficient $q+1-\rho_q$ is shown in Fig. \ref{fig:c1},
where the coefficient $q$ is included for comparison.

The proofs of Theorems \ref{the:fix}, \ref{the:high},
\ref{the:off}, and \ref{the:c1}, and
Corollary \ref{cor:high-exp} are given in
Section \ref{sec:proof}.

\section{Proofs of the Main Results}\label{sec:proof}

In Section \ref{sec:main}, we stated the main results of
this paper. We now provide the corresponding proofs. Recall that the $q$-ary differential VT codes
$\mathrm{DiffVT}_a(n;q)$ form a partition of
$\Sigma_q^n$, where $a\in[0,qn-1]$.
By the pigeonhole principle and Lemma \ref{lem:diffVT},
every integer $M$ satisfying $1\leq M\leq q^n/(qn)$
is feasible. For the code sizes in Theorems \ref{the:fix} and
\ref{the:high}, we have $qnM_n/q^n\to0$.
Moreover, for $M_{n,c}$ in Theorem \ref{the:off}
with $c\geq1$, we have $\frac{qnM_{n,c}}{q^n}\leq q^{1-c}\leq1.$
Therefore, the code sizes in Theorems \ref{the:fix}
and \ref{the:high}, and $M_{n,c}$ with $c\geq1$,
are feasible for all sufficiently large $n$.

We next establish an asymptotic estimate of
$\Prob\{Z_n\leq T\}$, which will be used repeatedly
in the proofs of the main results.
For this purpose, we first state the following
auxiliary lemma.
\begin{lemma}\cite[Theorem 1.2]{SiripraparatNeammanee2021}
\label{lem:loc}
Let $q\geq2$ and $\delta\in(0,1/q]$ be fixed.
Let $X_1,\ldots,X_n$ be i.i.d.\ random variables
taking values in $[q]$ with
$\Prob\{X_i=j\}\geq\delta$ for every $j\in[q]$.
Write $\mu\triangleq\E[X_i]$ and
$v\triangleq\Var(X_i)>0$.
There exists a constant $C_{q,\delta}>0$ such that,
for every integer $k$, we have
\begin{align*}
\left\lvert
\Prob\left\{\sum_{i=1}^nX_i=k\right\}
-\frac{\exp\bigl(-(k-n\mu)^2/(2nv)\bigr)}
{\sqrt{2\pi nv}}
\right\rvert
&\leq\frac{C_{q,\delta}}{n}.
\end{align*}
For fixed $q$ and $\delta$, the same constant
$C_{q,\delta}$ can be used regardless of the values of
$\Prob\{X_i=j\}$, provided that
$\Prob\{X_i=j\}\geq\delta$ for every $j\in[q]$.
This constant is also independent of $n$ and $k$.
\end{lemma}

\begin{proof}
Write $p_j\triangleq\Prob\{X_1=j\}$. Since $X_1\in[q]$,
we have $1\leq\E[\lvert X_1\rvert^3]\leq q^3$.
Moreover, $p_j\geq\delta$ gives
\begin{align*}
v
&=\frac12\sum_{i,j=1}^q p_ip_j(i-j)^2
\geq p_1p_2\geq\delta^2,
\quad
2n\sum_{j=1}^{q-1}p_jp_{j+1}
\geq2n(q-1)\delta^2.
\end{align*}
From \cite[Theorem 1.2]{SiripraparatNeammanee2021},
the absolute error in the statement is therefore
$O_{q,\delta}(n^{-2/3}e^{-c_{q,\delta}n^{1/3}}+n^{-1})
=O_{q,\delta}(n^{-1})$ for some $c_{q,\delta}>0$.
The constants depend only on $q$ and $\delta$,
which proves the required uniformity.
\end{proof}

We next apply Lemma \ref{lem:loc} to the auxiliary
random variables $U_i^\theta$ to obtain the following
estimate of $\Prob\{Z_n\leq T\}$.

\begin{lemma}\label{lem:tail}
Let $q\geq2$ be fixed and let
$Z_n=U_1+\cdots+U_n$, where $U_1,\ldots,U_n$ are independent
and uniformly distributed over $[q]$.

\begin{enumerate}
\item[(i)]
Let $J\subset(1,\mu_q)$ be a fixed compact interval.
If $T$ is an integer satisfying $T/n\in J$, then, as
$n\to\infty$, we have
\begin{align*}
\Prob\{Z_n\leq T\}
&=
\frac{e^{-nI_q(T/n)}}
{\sqrt{2\pi n\psi_q''(\theta_q(T/n))}}
\times
\frac{1+O_{q,J}(n^{-1/2})}
{1-e^{\theta_q(T/n)}},
\end{align*}
uniformly over all such $T$.

\item[(ii)]
Let $T_n$ be an integer.
If $T_n/n<\mu_q$ for all sufficiently large $n$,
$T_n/n\to\mu_q$, and
$n(\mu_q-T_n/n)^2\to\infty$, then, as $n\to\infty$, we have
\begin{align*}
\Prob\{Z_n\leq T_n\}
&=
\frac{e^{-nI_q(T_n/n)}}
{\sqrt{2\pi n\psi_q''(\theta_q(T_n/n))}}
\times
\frac{1+o_q(1)}
{1-e^{\theta_q(T_n/n)}}.
\end{align*}
\end{enumerate}
\end{lemma}

\begin{proof}
For part (i), let $J\subset(1,\mu_q)$ be the fixed
compact interval in the statement.
Let $T$ be any integer satisfying $T/n\in J$, and define
$\theta\triangleq\theta_q(T/n)$,
$v_\theta\triangleq\psi_q''(\theta)$, and
$Z_n^\theta\triangleq\sum_{i=1}^nU_i^\theta$.
Recall that $U_1^\theta,\ldots,U_n^\theta$ are i.i.d.\
random variables taking values in $[q]$, with
$\Prob\{U_i^\theta=j\}=p_\theta(j)
=e^{\theta j}/\sum_{a=1}^q e^{\theta a}$
for every $j\in[q]$.
Moreover $\theta<0$, $\E[U_i^\theta]=T/n$, and
$\Var(U_i^\theta)=v_\theta>0$.

By the definitions of $p_\theta$ and $\psi_q$, we have
$p_\theta(j)=q^{-1}e^{\theta j-\psi_q(\theta)}$
for every $j\in[q]$.
Then, for every integer $s$, independence gives
\begin{align*}
\Prob\left\{Z_n^\theta=s\right\}
&=
\sum_{\substack{
u_1,\ldots,u_n\in[q]\\
\sum_{i=1}^n u_i=s}}
\prod_{i=1}^n p_\theta(u_i)=
e^{\theta s-n\psi_q(\theta)}
\sum_{\substack{
u_1,\ldots,u_n\in[q]\\
\sum_{i=1}^n u_i=s}}
q^{-n}=
e^{\theta s-n\psi_q(\theta)}
\Prob\left\{Z_n=s\right\},
\end{align*}
where the last equality follows because $U_1,\ldots,U_n$
are i.i.d.\ uniform random variables over $[q]$.
Since $I_q(T/n)=\theta T/n-\psi_q(\theta)$, it follows that
\begin{align}
\Prob\left\{Z_n\leq T\right\}
&=\sum_{k=0}^{\infty}
e^{-\theta(T-k)+n\psi_q(\theta)}
\Prob\left\{Z_n^\theta=T-k\right\}
\notag\\
&=e^{-nI_q(T/n)}
\sum_{k=0}^{\infty}e^{\theta k}
\Prob\left\{Z_n^\theta=T-k\right\}.
\label{eq:tsum}
\end{align}
The summands vanish when $k>T-n$ since $Z_n^\theta\geq n$.

We next estimate
$\Prob\{Z_n^\theta=T-k\}$.
The function $\theta_q$ is continuous as the inverse of
the continuous and strictly increasing function $\psi_q'$, so the set
$\Theta\triangleq\{\theta_q(u):u\in J\}$
is a compact interval contained in $(-\infty,0)$.
The continuity and positivity of $p_\theta(j)$ and
$v_\theta$ imply
$\min_{\theta\in\Theta,\,j\in[q]}p_\theta(j)>0$
and
$0<\min_{\theta\in\Theta}v_\theta
\leq\max_{\theta\in\Theta}v_\theta<\infty$.
Thus, we can apply Lemma \ref{lem:loc} to the
i.i.d.\ random variables $U_1^\theta,\ldots,U_n^\theta$.
Their sum has mean $T$ and variance $nv_\theta$, so
\begin{align}
\Prob\left\{Z_n^\theta=T-k\right\}
&=
\frac{e^{-k^2/(2nv_\theta)}}{\sqrt{2\pi nv_\theta}}
+O_{q,\Theta}(n^{-1}),
\label{eq:tmass}
\end{align}
uniformly over $k\geq0$ and the integers $T$ under
consideration. Then, from equations \eqref{eq:tsum} and \eqref{eq:tmass},
we obtain
\begin{align}
\Prob\left\{Z_n\leq T\right\}
&=e^{-nI_q(T/n)}
\sum_{k=0}^{\infty}e^{\theta k}
\left[
\frac{e^{-k^2/(2nv_\theta)}}{\sqrt{2\pi nv_\theta}}
+O_{q,\Theta}(n^{-1})
\right]\notag\\
&=
\frac{e^{-nI_q(T/n)}}{\sqrt{2\pi nv_\theta}}
\sum_{k=0}^{\infty}e^{\theta k-k^2/(2nv_\theta)}
+O_{q,\Theta}\left(
\frac{e^{-nI_q(T/n)}}{n(1-e^\theta)}
\right),
\label{eq:tloc}
\end{align}
where the last term follows from
$\sum_{k=0}^{\infty}e^{\theta k}=1/(1-e^\theta)$.

It remains to estimate the sum in equation \eqref{eq:tloc}.
From the geometric-series identity, it can be easily
checked that
$\sum_{k=0}^{\infty}k^2e^{\theta k}
=e^\theta(1+e^\theta)/(1-e^\theta)^3$.
Using $0\leq1-e^{-x}\leq x$ for $x\geq0$, we have
\begin{align*}
0\leq
\frac1{1-e^\theta}
-\sum_{k=0}^{\infty}e^{\theta k-k^2/(2nv_\theta)}
&=
\sum_{k=0}^{\infty}e^{\theta k}
\left(1-e^{-k^2/(2nv_\theta)}\right)\leq
\frac1{2nv_\theta}
\sum_{k=0}^{\infty}k^2e^{\theta k}=
\frac{e^\theta(1+e^\theta)}
{2nv_\theta(1-e^\theta)^3}.
\end{align*}
Since $0<e^\theta<1$ and $v_\theta$ has a positive
lower bound on $\Theta$, we obtain
\begin{align}
\sum_{k=0}^{\infty}e^{\theta k-k^2/(2nv_\theta)}
&=
\frac1{1-e^\theta}
+O_{q,\Theta}\left(
\frac1{n(1-e^\theta)^3}
\right).
\label{eq:gsum}
\end{align}
Consequently, from equations \eqref{eq:tloc} and
\eqref{eq:gsum}, we have
\begin{align}
\Prob\left\{Z_n\leq T\right\}
&=
\frac{e^{-nI_q(T/n)}}
{(1-e^\theta)\sqrt{2\pi nv_\theta}}
+O_{q,\Theta}\left(
\frac{e^{-nI_q(T/n)}}{n^{3/2}(1-e^\theta)^3}
+\frac{e^{-nI_q(T/n)}}{n(1-e^\theta)}
\right)\notag\\
&=
\frac{e^{-nI_q(T/n)}}
{(1-e^\theta)\sqrt{2\pi n\psi_q''(\theta)}}
\left[
1+O_{q,\Theta}\left(
n^{-1/2}+\frac1{n(1-e^\theta)^2}
\right)
\right].
\label{eq:tall}
\end{align}
The argument leading to equation \eqref{eq:tall} applies
to any fixed compact interval $\Theta\subset(-\infty,0]$,
provided that $\theta=\theta_q(T/n)\in\Theta$ and $\theta<0$.

For the choice $\Theta=\{\theta_q(u):u\in J\}$,
we have $\min_{\theta\in\Theta}(1-e^\theta)>0$.
Therefore, the relative error in equation \eqref{eq:tall}
is $O_{q,J}(n^{-1/2})$, which proves part (i).

For part (ii), we have $T_n/n\in(1,\mu_q)$ for all
sufficiently large $n$.
By Lemma \ref{lem:tilt}(ii),
$\theta_q(T_n/n)\to0$, so we can take $\Theta=[-1,0]$
in equation \eqref{eq:tall}.
From the mean value theorem and the continuity of
$\psi_q''$, we obtain
\begin{align*}
\mu_q-\frac{T_n}{n}
&=
\psi_q'(0)-\psi_q'(\theta_q(T_n/n))
=
\sigma_q^2\lvert\theta_q(T_n/n)\rvert(1+o(1)).
\end{align*}
Moreover, Taylor's formula gives
$1-e^{\theta_q(T_n/n)}
=\lvert\theta_q(T_n/n)\rvert(1+o(1))$.
Together with the assumption
$n(\mu_q-T_n/n)^2\to\infty$, this implies
\begin{align*}
n(1-e^{\theta_q(T_n/n)})^2
&=
\frac{n(\mu_q-T_n/n)^2}{\sigma_q^4}(1+o(1))
\longrightarrow\infty.
\end{align*}
Hence, when $T=T_n$, the term
$O_{q,\Theta}\bigl(n^{-1/2}
+1/[n(1-e^{\theta_q(T_n/n)})^2]\bigr)$
in equation \eqref{eq:tall} is $o_q(1)$,
which proves part (ii).
\end{proof}

For later use, we give an equivalent form of part (ii).
Under its assumptions, $\theta_q(T_n/n)\to0$.
Since $\psi_q''$ is continuous and
$\psi_q''(0)=\sigma_q^2$, we have
$\psi_q''(\theta_q(T_n/n))\to\sigma_q^2$.
Then, by Taylor's formula and Lemma \ref{lem:rate}(i),
we obtain
\begin{align*}
I_q(T_n/n)
&=\theta_q(T_n/n)\psi_q'(\theta_q(T_n/n))
-\psi_q(\theta_q(T_n/n))
=\frac{\sigma_q^2}{2}\theta_q(T_n/n)^2(1+o(1)).
\end{align*}
Together with
$1-e^{\theta_q(T_n/n)}\sim\lvert\theta_q(T_n/n)\rvert$, we have
\begin{align*}
(1-e^{\theta_q(T_n/n)})
\sqrt{2\pi n\psi_q''(\theta_q(T_n/n))}
&=\sqrt{4\pi nI_q(T_n/n)}(1+o(1)).
\end{align*}
Consequently,
\begin{align}
\ln\Prob\{Z_n\leq T_n\}
&=-nI_q(T_n/n)
-\frac12\ln\bigl(4\pi nI_q(T_n/n)\bigr)+o_q(1).
\label{eq:tail-log}
\end{align}

\begin{remark}\label{rem:tail}
From Lemma \ref{lem:tail}, we see that $I_q(T/n)$
determines the exponential term $e^{-nI_q(T/n)}$
in the probability estimate.
The other factors involve $\theta_q(T/n)$ and
$\psi_q''(\theta_q(T/n))
=\Var(U_i^{\theta_q(T/n)})$.
For fixed $q$ and $\gamma\in(1,\mu_q)$,
$\theta_q(\gamma)$ and $\psi_q''(\theta_q(\gamma))$
are constants independent of $n$.
Moreover, if $T_n/n\to\gamma$, then
$\theta_q(T_n/n)\to\theta_q(\gamma)$ and
$\psi_q''(\theta_q(T_n/n))\to\psi_q''(\theta_q(\gamma))$
by continuity.
Together with
$L_q(n,T)/q^n=\Prob\{Z_n\leq T\}$, these relations
allow us to estimate $L_q(n,T)$.
\end{remark}

\subsection{Proof of Theorem \ref{the:fix}}

The upper and lower bounds in Theorem \ref{the:bound}
involve $L_q(n,T)$ and $L_q(n-1,T)$, respectively.
We first compare $T(L_q(n,T),qK)$ and
$T(L_q(n-1,T),K)$.
Recall that
$T(L_q(n,T),K)=\min\{t\geq n:L_q(n,t)\geq K\}$.

\begin{lemma}\label{lem:shift}
For integers $q\geq2$, $n\geq2$, and $1\leq K\leq q^{n-1}$,
we have
\begin{align*}
T(L_q(n-1,T),K)+1
&\leq T(L_q(n,T),qK)
\leq T(L_q(n-1,T),K)+q.
\end{align*}
\end{lemma}

\begin{proof}
Let $t\triangleq T(L_q(n-1,T),K)$.
Since $Z_n=Z_{n-1}+U_n$ and $1\leq U_n\leq q$, we have
\begin{gather*}
\Prob\{Z_n\leq t\}
\leq\Prob\{Z_{n-1}\leq t-1\}<\frac{K}{q^{n-1}},\\
\Prob\{Z_n\leq t+q\}
\geq\Prob\{Z_{n-1}\leq t\}\geq\frac{K}{q^{n-1}}.
\end{gather*}
Since $L_q(n,t)/q^n=\Prob\{Z_n\leq t\}$, we have $L_q(n,t)<qK\leq L_q(n,t+q).$
By Definition \ref{def:inv} and the monotonicity of
$L_q(n,T)$, we obtain
$t<T(L_q(n,T),qK)\leq t+q$.
Since both $t$ and $T(L_q(n,T),qK)$ are integers and
$t=T(L_q(n-1,T),K)$, it follows that
\begin{gather*}
T(L_q(n-1,T),K)+1
\leq T(L_q(n,T),qK)
\leq T(L_q(n-1,T),K)+q,
\end{gather*}
which proves the result.
\end{proof}

Next, we apply Lemma \ref{lem:tail}(i) to estimate
$T(L_q(n,T),K_n)$ when $K_n=Bq^{Rn}n^a(1+o(1))$ and obtain the following result.

\begin{lemma}\label{lem:finv}
Let $q\geq2$, $R\in(0,1)$, $B>0$, and $a\in\mathbb R$
be fixed. Let $\gamma\in(1,\mu_q)$ be the unique solution
of $I_q(\gamma)=(1-R)\ln q$, and define
$\theta\triangleq I_q'(\gamma)=\theta_q(\gamma)<0$.
If $K_n$ is a positive integer satisfying
$K_n=Bq^{Rn}n^a(1+o(1))$, then, as $n\to\infty$, we have
\begin{align*}
T(L_q(n,T),K_n)
&=n\gamma+\frac{a+1/2}{\lvert\theta\rvert}\ln n
+O_{q,R,B,a}(1).
\end{align*}
\end{lemma}

\begin{proof}
Let $t_n\triangleq T(L_q(n,T),K_n)$. By the definition of $t_n$, we have
$L_q(n,t_n-1)<K_n\leq L_q(n,t_n)$. Moreover,
for all sufficiently large $n$, we have $1<K_n<q^n$,
so $n<t_n\leq qn$ and $\ln K_n=Rn\ln q+a\ln n+\ln B+o(1).$
From Lemma \ref{lem:tail}(i), we know that for any
fixed compact interval $J\subset(1,\mu_q)$,
\begin{align}
\ln L_q(n,T)
&=n\ln q-nI_q(T/n)-\frac12\ln n+O_{q,J}(1)
\label{eq:flog}
\end{align}
uniformly over integers $T$ with $T/n\in J$.
For every sufficiently small fixed $\eps>0$, we obtain
from equation \eqref{eq:flog} and
$I_q(\gamma)=(1-R)\ln q$ that
\begin{gather*}
\lim_{n\to\infty}\frac1n
\ln\frac{L_q(n,\lfloor n(\gamma-\eps)\rfloor)}{K_n}
=
I_q(\gamma)-I_q(\gamma-\eps)<0,\\
\lim_{n\to\infty}\frac1n
\ln\frac{L_q(n,\lceil n(\gamma+\eps)\rceil)}{K_n}
=
I_q(\gamma)-I_q(\gamma+\eps)>0,
\end{gather*}
where the inequalities follow from the strict decrease
of $I_q$.
Consequently, for all sufficiently large $n$, we have
\begin{gather*}
L_q\bigl(n,\lfloor n(\gamma-\eps)\rfloor\bigr)
<K_n\leq
L_q\bigl(n,\lceil n(\gamma+\eps)\rceil\bigr).
\end{gather*}
Consequently, $t_n/n\to\gamma$.

We next determine $t_n-n\gamma$.
Since $t_n/n\to\gamma\in(1,\mu_q)$, both $t_n/n$ and
$(t_n-1)/n$ lie in a fixed compact subinterval of
$(1,\mu_q)$ for all sufficiently large $n$.
From equation \eqref{eq:flog}, we obtain
\begin{gather*}
0\leq\ln L_q(n,t_n)-\ln L_q(n,t_n-1)
=n\bigl(I_q((t_n-1)/n)-I_q(t_n/n)\bigr)+O(1)
=O(1),
\end{gather*}
where the last equality follows from the mean value theorem
and the boundedness of $I_q'$ on this interval.
Since
$L_q(n,t_n-1)<K_n\leq L_q(n,t_n)$, we have
\begin{gather*}
0\leq\ln L_q(n,t_n)-\ln K_n
<\ln L_q(n,t_n)-\ln L_q(n,t_n-1)
=O(1).
\end{gather*}
Hence, $\ln L_q(n,t_n)=\ln K_n+O(1)$.
Together with equation \eqref{eq:flog},
$\ln K_n=Rn\ln q+a\ln n+\ln B+o(1)$, and
$I_q(\gamma)=(1-R)\ln q$, we obtain $n\bigl(I_q(t_n/n)-I_q(\gamma)\bigr)
=-\left(a+\frac12\right)\ln n+O(1).$

Let $\Delta_n\triangleq t_n-n\gamma$.
By the mean value theorem, we have $n\bigl(I_q(t_n/n)-I_q(\gamma)\bigr)
=I_q'(\xi_n)\Delta_n,$
where $\xi_n$ lies between $\gamma$ and $t_n/n$.
Since $\xi_n\to\gamma$ and
$I_q'(\xi_n)\to I_q'(\gamma)=\theta\neq0$,
it follows that $\Delta_n=O(\ln n)$.
Moreover, from Lemma \ref{lem:rate}(i), we know that
$I_q''$ is continuous and bounded near $\gamma$.
Then, by Taylor's formula and $I_q'(\gamma)=\theta$, we have
\begin{gather*}
n\bigl(I_q(t_n/n)-I_q(\gamma)\bigr)
=
\theta\Delta_n
+\frac12 I_q''(\eta_n)\frac{\Delta_n^2}{n}
=
\theta\Delta_n+O\left(\frac{\Delta_n^2}{n}\right),
\end{gather*}
where $\eta_n$ lies between $\gamma$ and $t_n/n$.
The last equality follows from $\eta_n\to\gamma$
and the boundedness of $I_q''$ near $\gamma$.
Together with
$n(I_q(t_n/n)-I_q(\gamma))
=-(a+1/2)\ln n+O(1)$, we obtain $\theta\Delta_n+O\left(\frac{\Delta_n^2}{n}\right)
=-\left(a+\frac12\right)\ln n+O(1).$
Since $\Delta_n^2/n=O((\ln n)^2/n)=o(1)$ and $\theta<0$,
we conclude that $t_n=n\gamma+\Delta_n
=n\gamma+\frac{a+1/2}{\lvert\theta\rvert}\ln n+O(1),$
which proves the result.
\end{proof}

For a single-deletion correcting code, the number of
distinct deletion outputs is the sum of its run counts.
The next lemma relates this sum to code size through
the differential vector.
We will use it to prove the lower bound on
$\tau_q^{\del}(n,M_n)$ in Theorem \ref{the:fix}.
Its second conclusion will also be used in the next
subsection.

\begin{lemma}\label{lem:avg}
Let $q\geq2$ be fixed and let $p\triangleq(q-1)/q$.
For a nonempty code $\C\subseteq\Sigma_q^n$, where $n\geq2$,
define
\begin{align*}
\overline r(\C)
&\triangleq\frac1{\lvert\C\rvert}
\sum_{\bm{x}\in\C}r(\bm{x}),
\quad
a\triangleq\frac{\overline r(\C)-1}{n-1}.
\end{align*}
Let $h(t)\triangleq-t\ln t-(1-t)\ln(1-t)$ denote binary
entropy in natural logarithms, with $0\ln0=0$.
Then
\begin{align}
\ln\lvert\C\rvert
&\leq\ln q+(n-1)\bigl(h(a)+a\ln(q-1)\bigr).
\label{eq:avg}
\end{align}
For prescribed sizes $M_n$, the following results
hold uniformly over all
$\C\subseteq\Sigma_q^n$ with $\lvert\C\rvert=M_n$.
\begin{enumerate}
\item[(i)]
If $\ln M_n/n\to R\ln q$ for some fixed $R\in(0,1)$,
then there is a constant $c_{q,R}>0$ such that
$\overline r(\C)\geq c_{q,R}n$ for all sufficiently large $n$.
\item[(ii)]
If $\ln M_n=n\ln q-o(n)$, then
$\overline r(\C)=pn+o(n)$.
\end{enumerate}
\end{lemma}

\begin{proof}
Let $\bm{X}$ be uniformly distributed over $\C$, and put
$\bm{Y}\triangleq\Diff(\bm{X})$ and
$a_i\triangleq\Prob\{Y_i\ne0\}$ for $i\in[n-1]$.
The differential vector determines $\bm{X}$ recursively
from its last coordinate, and $r(\bm{X})=1+\sum_{i=1}^{n-1}\ind_{\{Y_i\ne0\}}$.
By the linearity of expectation, we have $\E[r(\bm{X})]
=
1+\sum_{i=1}^{n-1}
\Prob\{Y_i\ne0\}
=
1+\sum_{i=1}^{n-1}a_i.$
Then, from the definition of $a$, we obtain $\frac1{n-1}\sum_{i=1}^{n-1}a_i
=
\frac{\E[r(\bm{X})]-1}{n-1}
=a.$
Write $H$ for entropy in natural logarithms.
For $i\in[n-1]$ with $a_i>0$, the chain rule for entropy
gives
\begin{gather*}
H(Y_i)
=h(a_i)+a_iH(Y_i\mid Y_i\neq0)
\leq h(a_i)+a_i\ln(q-1),
\end{gather*}
where the inequality follows because $Y_i$ has at most
$q-1$ possible values when $Y_i\neq0$.
The same bound holds when $a_i=0$, since $Y_i=0$
with probability one.
Next, by the chain rule and the fact that conditioning
does not increase entropy, we have
\begin{gather*}
H(\bm{Y})
=H(Y_1)+\sum_{i=2}^{n}H(Y_i\mid Y_1,\ldots,Y_{i-1})
\leq\sum_{i=1}^{n}H(Y_i).
\end{gather*}
Together with $H(Y_n)\leq\ln q$, we obtain $\ln\lvert\C\rvert=H(\bm{Y})
\leq\ln q+\sum_{i=1}^{n-1}
\bigl(h(a_i)+a_i\ln(q-1)\bigr).$
Moreover, by the concavity of $h$ and Jensen's inequality,
\begin{gather*}
\sum_{i=1}^{n-1}h(a_i)
\leq(n-1)h\left(\frac1{n-1}\sum_{i=1}^{n-1}a_i\right)
=(n-1)h(a).
\end{gather*}
Since $\sum_{i=1}^{n-1}a_i=(n-1)a$, we conclude that $\ln\lvert\C\rvert
\leq\ln q+(n-1)\bigl(h(a)+a\ln(q-1)\bigr).$
This proves equation (\ref{eq:avg}).

Set $f(t)\triangleq h(t)+t\ln(q-1)$ for $t\in[0,1]$.
The function is continuous, with $f(0)=0$, and
$f'(t)=\ln((q-1)(1-t)/t)$ for $t\in(0,1)$.
Thus, $f$ is increasing on $[0,p]$ and has the unique
maximum $f(p)=\ln q$.
For part (i), choose $\delta\in(0,p)$ such that
$f(\delta)<R\ln q/2$. By equation \eqref{eq:avg}, we have $f(a)\geq\frac{\ln M_n-\ln q}{n-1}\longrightarrow R\ln q.$
It follows that $a>\delta$ for all sufficiently large $n$,
so $\overline r(\C)=1+(n-1)a\geq1+(n-1)\delta
=\delta n+(1-\delta)
\geq\delta n.$
We may therefore take $c_{q,R}=\delta$.
For part (ii), the same bound yields
\begin{align*}
\ln q\geq f(a)
&\geq\ln q-\frac{n\ln q-\ln M_n}{n-1}
=\ln q-o(1).
\end{align*}
The unique maximum of $f$ on the compact interval
$[0,1]$ implies $a\to p$, hence
$\overline r(\C)=pn+o(n)$.
Both arguments depend on the code only through $M_n$,
which proves the asserted uniformity.
\end{proof}

We now prove Theorem \ref{the:fix} by using
Lemmas \ref{lem:shift}, \ref{lem:finv}, and \ref{lem:avg}.

\begin{proof}[Proof of Theorem \ref{the:fix}]
All implicit constants below may depend on $q,R,A$.
Since $M_n=Aq^{Rn}(1+o(1))$, we apply
Lemma \ref{lem:finv} with $B=A$ and $a=0$.
It then follows from Theorem \ref{the:bound}(i) that
\begin{align*}
\tau_q(n,M_n)
&=T(L_q(n,T),M_n)
=n\gamma_{q,R}
+\frac{1}{2\lvert\theta_{q,R}\rvert}\ln n+O(1).
\end{align*}

For single-deletion correction, we have
$qnM_n=qAq^{Rn}n(1+o(1))$.
By Theorem \ref{the:bound}(ii) and Lemma \ref{lem:finv}
with $B=qA$ and $a=1$, we obtain
\begin{align*}
\tau_q^{\del}(n,M_n)
&\leq T(L_q(n,T),qnM_n)
=n\gamma_{q,R}
+\frac{3}{2\lvert\theta_{q,R}\rvert}\ln n+O(1).
\end{align*}

For the lower bound, consider any reference
$\S\in\Sigma_q^T$ and single-deletion correcting code
$\C\subseteq\Sub_n(\S)$ with $\lvert\C\rvert=M_n$.
From Corollaries \ref{cor:max-ref} and \ref{cor:del-up}
and Lemma \ref{lem:avg}(i), we have
\begin{align*}
L_q(n-1,T)
&\geq\lvert\Sub_{n-1}(\S)\rvert
\geq\sum_{\bm{x}\in\C}r(\bm{x})
\geq c_{q,R}nM_n.
\end{align*}
For all sufficiently large $n$, we have
$1\leq\lceil c_{q,R}nM_n\rceil\leq q^{n-1}$ and
$q\lceil c_{q,R}nM_n\rceil
=qc_{q,R}Aq^{Rn}n(1+o(1))$.
We apply Lemma \ref{lem:finv} with $B=qc_{q,R}A$
and $a=1$. Together with Lemma \ref{lem:shift},
we obtain
\begin{gather*}
T
\geq T\bigl(L_q(n-1,T),\lceil c_{q,R}nM_n\rceil\bigr)
\geq T\bigl(L_q(n,T),q\lceil c_{q,R}nM_n\rceil\bigr)-q
=n\gamma_{q,R}
+\frac{3}{2\lvert\theta_{q,R}\rvert}\ln n+O(1).
\end{gather*}
By the definition of $\tau_q^{\del}(n,M_n)$ and the
upper bound, we obtain the stated expansion.
\end{proof}

\subsection{Proof of Theorem \ref{the:high}}

We now prove Theorem \ref{the:high} by using
Theorem \ref{the:bound}, Lemma \ref{lem:tail}(ii),
and Lemma \ref{lem:avg}.
The following lemma estimates $T(L_q(n-\ell,T),K_n)$
for $\ell\in\{0,1\}$ when
$K_n/q^{n-\ell}=a q^{-s_n}(1+o(1))$
for a fixed $a>0$.

\begin{lemma}\label{lem:hinv}
Let $q\geq2$, $a>0$, and $\ell\in\{0,1\}$ be fixed.
Suppose that $s_n\to\infty$ and $s_n=o(n)$.
Let $\gamma_n\in(1,\mu_q)$ be the unique solution of
$I_q(\gamma_n)=s_n\ln q/n$ for all sufficiently large $n$,
and define $\theta_n\triangleq I_q'(\gamma_n)<0$.
If $K_n$ is a positive integer satisfying
$K_n/q^{n-\ell}=aq^{-s_n}(1+o(1))$, then
\begin{align*}
T(L_q(n-\ell,T),K_n)
&=n\gamma_n+
\frac{\frac12\ln(4\pi s_n\ln q)+\ln a}{\lvert\theta_n\rvert}
+o\left(\frac1{\lvert\theta_n\rvert}\right).
\end{align*}
\end{lemma}

The proof follows the same approach as that of
Lemma \ref{lem:finv}, using Lemma \ref{lem:tail}(ii)
to estimate the probability.
The complete proof is given in Appendix \ref{app:hinv}.

\begin{proof}[Proof of Theorem \ref{the:high}]
Since $s_n=o(n)$, we have
$M_n=q^{n-s_n}n^{-1}(1+o(1))$.
Then, $\frac{qnM_n}{q^n}
=q\cdot q^{-s_n}(1+o(1))\to0.$
By Theorem \ref{the:bound}(ii) and
Lemma \ref{lem:hinv} with $\ell=0$ and $a=q$,
we obtain
\begin{align*}
\tau_q^{\del}(n,M_n)
&\leq T(L_q(n,T),qnM_n)=n\gamma_n+
\frac{\frac12\ln(4\pi s_n\ln q)+\ln q}
{\lvert\theta_n\rvert}
+o_q\left(\frac1{\lvert\theta_n\rvert}\right).
\end{align*}

For the lower bound, let $p\triangleq(q-1)/q$.
From the expression for $M_n$, we have
$\ln M_n=n\ln q-s_n\ln q-\ln n+o(1)
=n\ln q-o(n)$.
By Lemma \ref{lem:avg}(ii), there exists
$\eps_n\geq0$ with $\eps_n\to0$, depending only on
$q$ and $M_n$, such that $\sum_{\bm{x}\in\C}r(\bm{x})\geq pnM_n(1-\eps_n)$
for every $\C\subseteq\Sigma_q^n$ with
$\lvert\C\rvert=M_n$.
We may take $\eps_n<1$ for all sufficiently large $n$.
For any $\S\in\Sigma_q^T$ and single-deletion correcting
code $\C\subseteq\Sub_n(\S)$ with $\lvert\C\rvert=M_n$,
we obtain from Corollaries \ref{cor:max-ref} and
\ref{cor:del-up} that
\begin{align*}
L_q(n-1,T)
&\geq\lvert\Sub_{n-1}(\S)\rvert
\geq\sum_{\bm{x}\in\C}r(\bm{x})
\geq pnM_n(1-\eps_n).
\end{align*}
Since $L_q(n-1,T)$ is an integer,
$L_q(n-1,T)\geq\lceil pnM_n(1-\eps_n)\rceil$.
By Definition \ref{def:inv} and the definition of
$\tau_q^{\del}(n,M_n)$, we have $\tau_q^{\del}(n,M_n)
\geq
T\bigl(L_q(n-1,T),\lceil pnM_n(1-\eps_n)\rceil\bigr).$
Moreover, $nM_n\to\infty$ and $\eps_n\to0$, so
\begin{align*}
\frac{\lceil pnM_n(1-\eps_n)\rceil}{q^{n-1}}
&=\frac{pnM_n}{q^{n-1}}(1+o(1))=\frac{pn}{q^{n-1}}
\frac{q^n}{nq^{s_n}}(1+o(1))\\
&=pq\cdot q^{-s_n}(1+o(1))
=(q-1)q^{-s_n}(1+o(1)).
\end{align*}
We can therefore apply Lemma \ref{lem:hinv}
with $\ell=1$ and $a=q-1$ to obtain
\begin{align*}
\tau_q^{\del}(n,M_n)
&\geq n\gamma_n+
\frac{\frac12\ln(4\pi s_n\ln q)+\ln(q-1)}
{\lvert\theta_n\rvert}
+o_q\left(\frac1{\lvert\theta_n\rvert}\right),
\end{align*}
which proves the result.
\end{proof}

We now prove Corollary \ref{cor:high-exp} using
Theorem \ref{the:high} and expansions at $\theta=0$.

\begin{proof}[Proof of Corollary \ref{cor:high-exp}]
By Lemma \ref{lem:tilt}(i) and Taylor's formula,
as $\theta\to0$ we have
$\psi_q(\theta)
=\mu_q\theta+\frac{\sigma_q^2}{2}\theta^2+O_q(\theta^4)$
and
$\psi_q'(\theta)
=\mu_q+\sigma_q^2\theta+O_q(\theta^3)$.
Together with $\theta_n<0,\gamma_n=\psi_q'(\theta_n)$ and
$I_q(\gamma_n)=\theta_n\psi_q'(\theta_n)-\psi_q(\theta_n)$,
these expansions give
$I_q(\gamma_n)=\frac{s_n\ln q}{n}
=\frac{\sigma_q^2}{2}\theta_n^2+O_q(\theta_n^4)$
and
$\mu_q-\gamma_n
=\sigma_q^2\lvert\theta_n\rvert
+O_q(\lvert\theta_n\rvert^3)$.
Since $I_q(\gamma_n)=s_n\ln q/n$, we obtain $\theta_n^2=O_q(s_n\ln q/n)$ and $\lvert\theta_n\rvert
=\frac1{\sigma_q}\sqrt{\frac{2s_n\ln q}{n}}
\left(1+O_q\left(\frac{s_n\ln q}{n}\right)\right).$
Consequently,
$n\gamma_n
=\mu_qn-\sigma_q\sqrt{2ns_n\ln q}
+O_q\left(\frac{(s_n\ln q)^{3/2}}{\sqrt n}\right)$
and
$\frac1{\lvert\theta_n\rvert}
=\sigma_q\sqrt{\frac{n}{2s_n\ln q}}
\left(1+O_q\left(\frac{s_n\ln q}{n}\right)\right)$.
The assumption $s_n=o(\sqrt n)$ implies
$\frac{(s_n\ln q)^{3/2}/\sqrt n}{\sqrt{n/(s_n\ln q)}}
=\frac{(s_n\ln q)^2}{n}\to0$
and
$\frac{s_n\ln q\,\ln(s_n\ln q)}{n}\to0$.
For $a\in\{q-1,q\}$, we have
$\frac12\ln(4\pi s_n\ln q)+\ln a
=O_q(\ln(s_n\ln q))$.
Hence,
\begin{align*}
O_q\left(\frac{(s_n\ln q)^{3/2}}{\sqrt n}\right)
&=
\sqrt{\frac{n}{s_n\ln q}}\,
O_q\left(\frac{(s_n\ln q)^2}{n}\right)
=o_q\left(\sqrt{\frac{n}{s_n\ln q}}\right),\\
\sqrt{\frac{n}{s_n\ln q}}\,
O_q\left(\frac{s_n\ln q}{n}\right)
O_q\left(\ln(s_n\ln q)\right)
&=
\sqrt{\frac{n}{s_n\ln q}}\,
O_q\left(\frac{s_n\ln q\,\ln(s_n\ln q)}{n}\right)
=o_q\left(\sqrt{\frac{n}{s_n\ln q}}\right).
\end{align*}
Together with
$o_q(1/\lvert\theta_n\rvert)
=o_q(\sqrt{n/(s_n\ln q)})$,
substitution into Theorem \ref{the:high} proves the result.
\end{proof}

\subsection{Proof of Theorem \ref{the:off}}

We first estimate $W_{q,m}(N)$ and then determine the
smallest $T$ for which $L_q(m,T)/q^m$ reaches a fixed
value in $(0,1)$.
Recall that
$W_{q,m}(N)=
\max_{\B\subseteq\Sigma_q^m,\,\lvert\B\rvert\leq N}
\sum_{\bm{y}\in\B}1/r(\bm{y})$.
The following lemma gives this estimate and the exact
value of $W_{q,m}(N)$ when $N=q^m$.

\begin{lemma}\label{lem:w}
Let $q\geq2$ be fixed and let $p\triangleq(q-1)/q$.
As $m\to\infty$, uniformly over integers
$0\leq N\leq q^m$, we have
\begin{align*}
W_{q,m}(N)
&=\frac{N}{pm}
+O_q\left(\frac{q^m}{m^{3/2}}\right).
\end{align*}
Moreover, for every integer $m\geq1$, $W_{q,m}(q^m)=\frac{q^m-1}{pm}.$
\end{lemma}

\begin{proof}
Let $\bm{X}$ be uniformly distributed over $\Sigma_q^m$
and define $J_m\triangleq r(\bm{X})$.
The map $\Diff$ is a bijection on $\Sigma_q^m$,
with inverse
$X_i=(\sum_{j=i}^m\Diff(\bm{X})_j)\bmod q$.
Thus, the coordinates of $\Diff(\bm{X})$ are independent
and uniform over $\Sigma_q$.
Since
$J_m=1+\sum_{i=1}^{m-1}\ind_{\{\Diff(\bm{X})_i\ne0\}}$,
the random variable $J_m-1$ has the binomial distribution
with parameters $m-1$ and $p$.
By the definition of expectation, we have
\begin{align*}
\E\left[\frac1{J_m}\right]
&=\sum_{j=0}^{m-1}
\frac1{j+1}\binom{m-1}{j}p^j(1-p)^{m-1-j}\\
&=\int_0^1
\sum_{j=0}^{m-1}
\binom{m-1}{j}(pt)^j(1-p)^{m-1-j}\,dt\\
&=\int_0^1(1-p+pt)^{m-1}\,dt=\frac{1-(1-p)^m}{pm},
\end{align*}
where the second equality follows from
$1/(j+1)=\int_0^1t^j\,dt$,
and the third equality follows from the binomial theorem.
In particular,
$W_{q,m}(q^m)=\sum_{\bm{y}\in\Sigma_q^m}\frac1{r(\bm{y})}=q^m\E[1/J_m]=(q^m-1)/(pm)$.
Together with $\E[J_m]=pm+1-p$, this identity yields
\begin{align*}
\E\left[\frac{(J_m-pm)^2}{J_m}\right]
&=\E[J_m]-2pm+(pm)^2\E\left[\frac1{J_m}\right]
=1-p-pm(1-p)^m\leq1-p.
\end{align*}
The Cauchy--Schwarz inequality therefore gives
\begin{align*}
\E\left[\left\lvert\frac1{J_m}-\frac1{pm}\right\rvert\right]
=\frac1{pm}\E\left[\frac{\lvert J_m-pm\rvert}{J_m}\right]\leq\frac1{pm}
\sqrt{\E\left[\frac{(J_m-pm)^2}{J_m}\right]
\E\left[\frac1{J_m}\right]}
\leq\frac{\sqrt{1-p}}{(pm)^{3/2}}.
\end{align*}
For any $\B\subseteq\Sigma_q^m$ with $\lvert\B\rvert=N$,
it follows that
\begin{align}
\left\lvert
\sum_{\bm{y}\in\B}\frac1{r(\bm{y})}-\frac{N}{pm}
\right\rvert
\leq
\sum_{\bm{y}\in\Sigma_q^m}
\left\lvert\frac1{r(\bm{y})}-\frac1{pm}\right\rvert=
q^m\E\left[
\left\lvert\frac1{J_m}-\frac1{pm}\right\rvert
\right]
=O_q\left(\frac{q^m}{m^{3/2}}\right).
\label{eq:werr}
\end{align}
The bound is independent of $\B$ and $N$.
Since $1/r(\bm{y})>0$, adding a sequence increases
the sum whenever $\lvert\B\rvert<N$.
Thus, the maximum defining $W_{q,m}(N)$ is attained
by a set with exactly $N$ sequences.
From equation \eqref{eq:werr}, we conclude that
\begin{align*}
W_{q,m}(N)
&=
\max_{\substack{
\B\subseteq\Sigma_q^m\\
\lvert\B\rvert=N}}
\sum_{\bm{y}\in\B}\frac1{r(\bm{y})}
=\frac{N}{pm}+O_q\left(\frac{q^m}{m^{3/2}}\right)
\end{align*}
uniformly over $0\leq N\leq q^m$.
\end{proof}

Recall that $\Phi$ denotes the cumulative distribution
function of the standard normal distribution.
Here, we need to estimate $L_q(m,T)/q^m$ when it
approaches a constant in $(0,1)$, whereas under the
assumptions of Lemma \ref{lem:tail}, this ratio tends
to zero.
Therefore, we use the following Berry--Esseen inequality.

\begin{lemma}\cite[Chapter V]{Petrov1975}\label{lem:be}
Let $V_1,\ldots,V_m$ be i.i.d.\ random variables with
$\E[V_i]=0$, $\Var(V_i)=\sigma^2>0$, and
$\rho\triangleq\E[\lvert V_i\rvert^3]<\infty$.
There exists an absolute constant $C>0$ such that
\begin{align*}
\sup_{x\in\mathbb R}
\left\lvert
\Prob\left\{
\frac{\sum_{i=1}^mV_i}{\sigma\sqrt m}\leq x
\right\}
-\Phi(x)
\right\rvert
&\leq\frac{C\rho}{\sigma^3\sqrt m}.
\end{align*}
\end{lemma}

Recall that $U_1,\ldots,U_m$ are i.i.d.\ uniform
random variables over $[q]$.
For $i\in[m]$, define $V_i\triangleq U_i-\mu_q$.
Then, $V_1,\ldots,V_m$ are i.i.d.\ with
$\E[V_i]=0$ and $\Var(V_i)=\sigma_q^2>0$.
Moreover, $\E[\lvert V_i\rvert^3]
=\frac1q\sum_{j=1}^q\lvert j-\mu_q\rvert^3
\leq\left(\frac{q-1}{2}\right)^3<\infty.$
Since $\sum_{i=1}^mV_i=Z_m-m\mu_q$, we obtain from
Lemma \ref{lem:be} and
$L_q(m,T)/q^m=\Prob\{Z_m\leq T\}$ that
\begin{align}
\frac{L_q(m,T)}{q^m}
&=\Prob\{Z_m\leq T\}
=\Phi\left(\frac{T-\mu_qm}{\sigma_q\sqrt m}\right)
+O_q(m^{-1/2}),
\label{eq:norm}
\end{align}
uniformly over integers $T\geq m$.
The following lemma converts these probability estimates
into synthesis-time estimates.

\begin{lemma}\label{lem:ninv}
Let $q\geq2$, $\alpha\in(0,1)$, and $C>0$ be fixed.
The following results hold as $m\to\infty$.
\begin{enumerate}
\item[(i)]
If $K_m$ is an integer satisfying
$\lvert K_m/q^m-\alpha\rvert\leq C/\sqrt m$, then
\begin{align*}
T(L_q(m,T),K_m)
&=\mu_qm+\sigma_q\Phi^{-1}(\alpha)\sqrt m
+O_{q,\alpha,C}(1).
\end{align*}
\item[(ii)]
If $T\geq m$ is an integer satisfying
$L_q(m,T)/q^m\geq\alpha-C/\sqrt m$, then
\begin{align*}
T
&\geq\mu_qm+\sigma_q\Phi^{-1}(\alpha)\sqrt m
-O_{q,\alpha,C}(1).
\end{align*}
\end{enumerate}
\end{lemma}

\begin{proof}
For part (i), let $t_m\triangleq T(L_q(m,T),K_m)$.
Since $\alpha\in(0,1)$ is fixed, we have
$1<K_m<q^m$ for all sufficiently large $m$.
In particular, $t_m>m$ because $L_q(m,m)=1$.
By Definition \ref{def:inv}, we have $\frac{L_q(m,t_m-1)}{q^m}
<\frac{K_m}{q^m}\leq\frac{L_q(m,t_m)}{q^m}.$
Furthermore, for every $x\in\mathbb R$, we have
$\Phi'(x)=e^{-x^2/2}/\sqrt{2\pi}\leq1/\sqrt{2\pi}$.
By the mean value theorem,
\begin{align*}
0
&\leq
\Phi\left(\frac{t_m-\mu_qm}{\sigma_q\sqrt m}\right)
-\Phi\left(\frac{t_m-1-\mu_qm}{\sigma_q\sqrt m}\right)\leq
\frac1{\sqrt{2\pi}}
\left(
\frac{t_m-\mu_qm}{\sigma_q\sqrt m}
-\frac{t_m-1-\mu_qm}{\sigma_q\sqrt m}
\right)=\frac1{\sigma_q\sqrt{2\pi m}}.
\end{align*}
Together with equation \eqref{eq:norm} and
$K_m/q^m=\alpha+O_C(m^{-1/2})$, this gives $\Phi\left(\frac{t_m-\mu_qm}{\sigma_q\sqrt m}\right)
=\alpha+O_{q,C}(m^{-1/2}).$
Since $(\Phi^{-1})'(x)=1/\Phi'(\Phi^{-1}(x))$
is continuous and bounded near $\alpha$, the mean
value theorem gives
\begin{align}
\Phi^{-1}(\alpha+\delta)
&=\Phi^{-1}(\alpha)+O_\alpha(\lvert\delta\rvert)
\label{eq:pinv}
\end{align}
as $\delta\to0$.
It follows that
$(t_m-\mu_qm)/(\sigma_q\sqrt m)
=\Phi^{-1}(\alpha)+O_{q,\alpha,C}(m^{-1/2})$,
which proves part (i).
For part (ii), equation \eqref{eq:norm} implies $\Phi\left(\frac{T-\mu_qm}{\sigma_q\sqrt m}\right)
\geq\alpha-\frac{C+C_q}{\sqrt m}$
for some constant $C_q>0$.
For all sufficiently large $m$, we have
$\alpha-(C+C_q)/\sqrt m\in(0,1)$.
By the monotonicity of $\Phi$ and equation \eqref{eq:pinv},
we obtain
\begin{align*}
\frac{T-\mu_qm}{\sigma_q\sqrt m}
&\geq
\Phi^{-1}\left(\alpha-\frac{C+C_q}{\sqrt m}\right)\geq
\Phi^{-1}(\alpha)-O_{q,\alpha,C}(m^{-1/2}).
\end{align*}
Consequently,
\begin{align*}
T
&\geq
\mu_qm+\sigma_q\Phi^{-1}(\alpha)\sqrt m
-\sigma_q\sqrt m\,O_{q,\alpha,C}(m^{-1/2})=
\mu_qm+\sigma_q\Phi^{-1}(\alpha)\sqrt m
-O_{q,\alpha,C}(1),
\end{align*}
which proves part (ii).
\end{proof}

We now prove Theorem \ref{the:off} by using
Theorem \ref{the:bound}(ii) and
Lemmas \ref{lem:w} and \ref{lem:ninv}.

\begin{proof}[Proof of Theorem \ref{the:off}]
Let $p\triangleq(q-1)/q$.
By the definition of $M_{n,c}$, we have
\begin{align}
M_{n,c}
&=\frac{q^n}{nq^c}+O(1),
\quad
\frac{nM_{n,c}}{q^n}
=q^{-c}+O(nq^{-n}).
\label{eq:offm}
\end{align}
We first prove part (i).
Suppose that $\C\subseteq\Sigma_q^n$ is a
single-deletion correcting code with
$\lvert\C\rvert=M_{n,c}$.
Since $\A_{q,qn}$ synthesizes every sequence in
$\Sigma_q^n$ and $\Sigma_q^{n-1}$, we have
$\C\subseteq\Sub_n(\A_{q,qn})$ and
$L_q(n-1,qn)=q^{n-1}$.
From Theorem \ref{the:bound}(ii) and the exact formula
in Lemma \ref{lem:w}, we obtain $M_{n,c}
\leq W_{q,n-1}(q^{n-1})
=\frac{q^n-q}{(q-1)(n-1)}.$
Together with equation \eqref{eq:offm}, this gives
\begin{align*}
(q-1)q^{-c}+O_q(nq^{-n})
&=\frac{(q-1)nM_{n,c}}{q^n}
\leq\frac{n}{n-1}(1-q^{1-n})
=1+O_q(n^{-1}).
\end{align*}
When $c<\log_q(q-1)$, we have $(q-1)q^{-c}>1$,
so no such code exists for all sufficiently large $n$.
This proves part (i).

For part (ii), assume $c>\log_q(q-1)$.
If $M_{n,c}$ is not feasible, the lower bound holds
because $\tau_q^{\del}(n,M_{n,c})=+\infty$.
Otherwise, consider any $\S\in\Sigma_q^T$ and
single-deletion correcting code
$\C\subseteq\Sub_n(\S)$ with
$\lvert\C\rvert=M_{n,c}$.
By Theorem \ref{the:bound}(ii) and Lemma \ref{lem:w},
we have
$M_{n,c}\leq W_{q,n-1}(L_q(n-1,T))
=\frac{L_q(n-1,T)}{p(n-1)}
+O_q\left(\frac{q^{n-1}}{(n-1)^{3/2}}\right)$.
Moreover, from equation \eqref{eq:offm}, we obtain
\begin{align*}
\frac{p(n-1)M_{n,c}}{q^{n-1}}
&=pq\frac{n-1}{n}
\left(q^{-c}+O(nq^{-n})\right)=(q-1)q^{-c}\left(1-\frac1n\right)
+O_q(nq^{-n}).
\end{align*}
Consequently, there exists a constant $C_{q,c}>0$
independent of $\S$ and $\C$ such that
\begin{align*}
\frac{L_q(n-1,T)}{q^{n-1}}
&\geq
\frac{p(n-1)M_{n,c}}{q^{n-1}}
-O_q((n-1)^{-1/2})\geq
(q-1)q^{-c}-C_{q,c}(n-1)^{-1/2}.
\end{align*}
Since $(q-1)q^{-c}\in(0,1)$, we obtain from
Lemma \ref{lem:ninv}(ii) with $\alpha=(q-1)q^{-c}$ that
\begin{align*}
T
&\geq
\mu_q(n-1)
+\sigma_q\Phi^{-1}((q-1)q^{-c})\sqrt{n-1}
-O_{q,c}(1)=
\mu_qn+\sigma_q\Phi^{-1}((q-1)q^{-c})\sqrt n
-O_{q,c}(1),
\end{align*}
where we used $\sqrt{n-1}=\sqrt n+O(n^{-1/2})$.
By the definition of $\tau_q^{\del}(n,M_{n,c})$,
we obtain part (ii).

Finally, assume $c>1$.
From equation \eqref{eq:offm}, we have $\frac{qnM_{n,c}}{q^n}
=q^{1-c}+O_q(nq^{-n})
=q^{1-c}+O_q(n^{-1/2}).$
Since $q^{1-c}\in(0,1)$, it follows from
Theorem \ref{the:bound}(ii) and
Lemma \ref{lem:ninv}(i) with $\alpha=q^{1-c}$ that
\begin{align*}
\tau_q^{\del}(n,M_{n,c})
&\leq T(L_q(n,T),qnM_{n,c})=\mu_qn+\sigma_q\Phi^{-1}(q^{1-c})\sqrt n
+O_{q,c}(1),
\end{align*}
which proves part (iii).
For fixed $c>1$, both normal quantiles are constants.
Combining parts (ii) and (iii), we obtain
\begin{align*}
\tau_q^{\del}(n,M_{n,c})
&=\mu_qn+O_{q,c}(\sqrt n),
\quad
\lim_{n\to\infty}
\frac{\tau_q^{\del}(n,M_{n,c})}{n}
=\mu_q.
\end{align*}
\end{proof}

\subsection{Proof of Theorem \ref{the:c1}}

To prove Theorem \ref{the:c1}, we construct a single-deletion correcting code with at least $M_{n,1}$
codewords that can be synthesized within the time
stated in the theorem.
For this construction, we first recall the following
result on binary single-deletion correcting codes
whose codewords have the same Hamming weight.

\begin{lemma}\cite[Lemma 4]{Cullina2012}\label{lem:cw}
Let $n\geq2$, $k\in[n-1]$, and
$m_k\triangleq\max(k,n-k)+1$.
For $a\in[0,m_k-1]$, the set
\begin{align*}
\left\{
\bm{x}\in\{0,1\}^n:
\sum_{i=1}^n x_i=k,\quad
\sum_{i=1}^n ix_i\equiv a\pmod{m_k}
\right\}
\end{align*}
is a single-deletion correcting code.
\end{lemma}

We next use Lemmas \ref{lem:diffVT} and \ref{lem:cw}
to construct a single-deletion correcting code in
$\Sub_n(\A_{q,T})$ and estimate its size.
The use of composition classes in this construction
follows \cite{Cullina2016}.

\begin{lemma}\label{lem:comp}
Let $q\geq3$ and $\delta>0$ be fixed.
For all sufficiently large $n$, if $T$ is an integer
satisfying $T\geq(3q/4+\delta)n$, then there exists
a single-deletion correcting code
$\C\subseteq\Sub_n(\A_{q,T})$ such that
\begin{align*}
\lvert\C\rvert
&\geq
\frac{L_q(n,T)+(1-O_{q,\delta}(n^{-1/2}))2^n}{qn}.
\end{align*}
\end{lemma}

\begin{proof}
Our construction mainly uses differential VT codes,
while the binary sequences are treated separately
using Lemma \ref{lem:cw} to obtain a better bound on
code size. The resulting subcodes are then combined
into a single-deletion correcting code so that we can obtain the desired lower bound on the code size. 

We first construct the code.
Set $\B\triangleq\Sub_n(\A_{q,T})$ and partition
$\B$ by the \emph{composition}
$\bm{k}=(k_0,\ldots,k_{q-1})$, where $k_j$ is the
number of occurrences of symbol $j$.
Denote the corresponding subsets by $\B_{\bm{k}}$.
If $k_2+\cdots+k_{q-1}>0$, we define
\begin{align*}
\C_{\bm{k}}
&\triangleq
\B_{\bm{k}}\cap\mathrm{DiffVT}_{a_{\bm{k}}}(n;q).
\end{align*}
Otherwise, write $\bm{k}=(n-k,k,0,\ldots,0)$
with $k\in[0,n]$ and let
$m_k\triangleq\max(k,n-k)+1$.
We take
\begin{align*}
\C_{\bm{k}}
&\triangleq
\left\{
\bm{x}\in\B_{\bm{k}}:
\sum_{i=1}^n ix_i\equiv a_{\bm{k}}\pmod{m_k}
\right\}.
\end{align*}
In both cases, $a_{\bm{k}}$ is chosen to maximize
$\lvert\C_{\bm{k}}\rvert$, with its range being
$[0,qn-1]$ and $[0,m_k-1]$, respectively.
In particular, $\C_{\bm{k}}=\B_{\bm{k}}$
when $k=0$ or $k=n$.
We define
\begin{align*}
\C
&\triangleq
\bigcup_{\sum_{j=0}^{q-1}jk_j\equiv b_0\pmod q}
\C_{\bm{k}},
\end{align*}
where $b_0\in\Sigma_q$ is chosen to maximize the
size of this union.

We next prove that $\C$ corrects one deletion.
By Lemmas \ref{lem:diffVT} and \ref{lem:cw},
each $\C_{\bm{k}}$ is a single-deletion correcting code
for $\bm{k}\notin\{(n,0,\ldots,0),(0,n,0,\ldots,0)\}$.
The same holds for the remaining two compositions,
since
$\lvert\C_{(n,0,\ldots,0)}\rvert\leq1$ and
$\lvert\C_{(0,n,0,\ldots,0)}\rvert\leq1$.
For any distinct $\bm{x},\bm{x}'\in\C$, suppose that
$\bm{y}\in\D_1(\bm{x})\cap\D_1(\bm{x}')$.
Let $u,v\in\Sigma_q$ be the respective deleted symbols in $\bm{x},\bm{x}'$.
By the definition of $\C$, we have
\begin{align*}
u+\sum_{i=1}^{n-1}y_i
&\equiv b_0
\equiv v+\sum_{i=1}^{n-1}y_i\pmod q.
\end{align*}
Thus, $u\equiv v\pmod q$, so $u=v$. Consequently, $\bm{x},\bm{x}'$ have the same composition $\bm{k}$, i.e., $\bm{x},\bm{x}'\in\C_{\bm{k}}$ for some $\bm{k}$.
Since $\bm{x}\neq\bm{x}'$ and $\C_{\bm{k}}$ can correct
one deletion, we have
$\D_1(\bm{x})\cap\D_1(\bm{x}')=\varnothing$, which is
a contradiction.
Hence, $\C$ is a single-deletion correcting code.

It remains to estimate $\lvert\C\rvert$.
For every $\bm{x}\in\B_{\bm{k}}$, by the definition of
$\Diff$, we have $\Diff(\bm{x})_n=x_n$ and
$\Diff(\bm{x})_i=x_i-x_{i+1}
+q\ind_{\{x_i<x_{i+1}\}}$ for $i\in[n-1]$.
Therefore,
\begin{align*}
\sum_{i=1}^n i\Diff(\bm{x})_i
&=\sum_{i=1}^{n-1}i(x_i-x_{i+1})+nx_n
+q\sum_{i=1}^{n-1}i\ind_{\{x_i<x_{i+1}\}}\\
&=\sum_{i=1}^n ix_i-\sum_{i=2}^n(i-1)x_i
+q\sum_{i=1}^{n-1}i\ind_{\{x_i<x_{i+1}\}}\\
&=\sum_{i=1}^n x_i
+q\sum_{i=1}^{n-1}i\ind_{\{x_i<x_{i+1}\}}=\sum_{j=0}^{q-1}jk_j
+q\sum_{i=1}^{n-1}i\ind_{\{x_i<x_{i+1}\}}.
\end{align*}
Thus, $\sum_{i=1}^n i\Diff(\bm{x})_i
\equiv\sum_{j=0}^{q-1}jk_j\pmod q$
for every $\bm{x}\in\B_{\bm{k}}$, and $\left\lvert
\left\{
a\in[0,qn-1]:
\B_{\bm{k}}\cap\mathrm{DiffVT}_a(n;q)\neq\varnothing
\right\}
\right\rvert
\leq n.$
By the pigeonhole principle and the choice of
$a_{\bm{k}}$, we have
$\lvert\C_{\bm{k}}\rvert\geq\lvert\B_{\bm{k}}\rvert/n$
when $k_2+\cdots+k_{q-1}>0$.
For $\bm{k}=(n-k,k,0,\ldots,0)$ with
$1\leq k\leq n-1$, the same principle gives
$\lvert\C_{\bm{k}}\rvert\geq\lvert\B_{\bm{k}}\rvert/m_k$.
This inequality also holds for $k=0,n$, since
$\C_{\bm{k}}=\B_{\bm{k}}$ and $m_k=n+1$.
Let $\bm{X}$ be uniformly distributed over $\{0,1\}^n$
and define $K\triangleq\sum_{i=1}^nX_i$.
Summing the bounds on $\lvert\C_{\bm{k}}\rvert$ gives
\begin{align}
\sum_{\bm{k}}\lvert\C_{\bm{k}}\rvert
&\geq\frac{\lvert\B\rvert}{n}
+\sum_{k=0}^n
\left(\frac1{m_k}-\frac1n\right)
\lvert\B_{(n-k,k,0,\ldots,0)}\rvert\notag\\
&=\frac{\lvert\B\rvert}{n}
+\frac{2^n}{n}\E\left[
\left(\frac{n}{m_K}-1\right)
\ind_{\{\bm{X}\in\B\}}
\right],
\label{eq:csum}
\end{align}
where the equality follows from
$\Prob\{K=k,\bm{X}\in\B\}
=2^{-n}\lvert\B_{(n-k,k,0,\ldots,0)}\rvert$
for $k\in[0,n]$.
Here $K$ follows a binomial distribution with
parameters $n$ and $1/2$, and
$m_K=n/2+\lvert K-n/2\rvert+1$.
By the Cauchy--Schwarz inequality,
$\E[\lvert K-n/2\rvert]\leq\sqrt{\Var(K)}=\sqrt n/2$.
Consequently,
\begin{align*}
\E\left[\left\lvert\frac{n}{m_K}-2\right\rvert\right]
&=\E\left[\frac{2(\lvert K-n/2\rvert+1)}{m_K}\right]\leq\frac4n\left(\E[\lvert K-n/2\rvert]+1\right)
\leq\frac4n\left(\frac{\sqrt n}{2}+1\right)
=O(n^{-1/2}).
\end{align*}

To estimate $\Prob\{\bm{X}\notin\B\}$, let
$H\triangleq\sum_{i=1}^{n-1}\ind_{\{X_i<X_{i+1}\}}$.
The indicators $\ind_{\{X_i<X_{i+1}\}}$ and
$\ind_{\{X_j<X_{j+1}\}}$ are independent when
$\lvert i-j\rvert>1$.
Thus, $\E[H]=(n-1)/4$ and $\Var(H)=O(n)$.
From the definition of the interval vector and
Lemma \ref{lem:ref-time}, we have
\begin{align*}
t_{\A_{q,qn}}(\bm{X})
&=\sum_{i=1}^n\Int(\bm{X})_i
=X_n+1+q(n-1-H).
\end{align*}
It follows that
$\E[t_{\A_{q,qn}}(\bm{X})]=3qn/4+O_q(1)$
and
\begin{align*}
\Var(t_{\A_{q,qn}}(\bm{X}))
&=\Var(X_n-qH)
\leq2\Var(X_n)+2q^2\Var(H)
=O_q(n).
\end{align*}
Since $T\geq(3q/4+\delta)n$, we have
$T-\E[t_{\A_{q,qn}}(\bm{X})]\geq\delta n/2$
for all sufficiently large $n$.
By Chebyshev's inequality,
\begin{align*}
\Prob\{\bm{X}\notin\B\}
&=\Prob\{t_{\A_{q,qn}}(\bm{X})>T\}\leq
\frac{\Var(t_{\A_{q,qn}}(\bm{X}))}
{(T-\E[t_{\A_{q,qn}}(\bm{X})])^2}
=O_{q,\delta}(n^{-1}).
\end{align*}
Moreover, $\lvert n/m_K-1\rvert\leq1$, so
\begin{align*}
\left\lvert
\E\left[
\left(\frac{n}{m_K}-1\right)
\ind_{\{\bm{X}\in\B\}}
\right]-1
\right\rvert
&\leq
\E\left[\left\lvert\frac{n}{m_K}-2\right\rvert\right]
+\Prob\{\bm{X}\notin\B\}=O_{q,\delta}(n^{-1/2}).
\end{align*}
Then, from equation \eqref{eq:csum} and
$\lvert\B\rvert=L_q(n,T)$, we obtain
\begin{align*}
\sum_{\bm{k}}\lvert\C_{\bm{k}}\rvert
&\geq
\frac{L_q(n,T)+(1-O_{q,\delta}(n^{-1/2}))2^n}{n}.
\end{align*}
For $b\in\Sigma_q$, define $\C_b\triangleq
\bigcup_{\sum_{j=0}^{q-1}jk_j\equiv b\pmod q}
\C_{\bm{k}}.$
The sets $\C_{\bm{k}}$ are pairwise disjoint,
and each composition determines exactly one residue
$b\equiv\sum_{j=0}^{q-1}jk_j\pmod q$.
By the choice of $b_0$, we obtain
\begin{align*}
\lvert\C\rvert
&=\lvert\C_{b_0}\rvert
=\max_{b\in\Sigma_q}\lvert\C_b\rvert\geq\frac1q\sum_{b\in\Sigma_q}\lvert\C_b\rvert=\frac1q\sum_{\bm{k}}\lvert\C_{\bm{k}}\rvert\geq
\frac{L_q(n,T)+(1-O_{q,\delta}(n^{-1/2}))2^n}{qn},
\end{align*}
which proves the result.
\end{proof}

We now apply Lemma \ref{lem:comp} to prove
Theorem \ref{the:c1}.

\begin{proof}[Proof of Theorem \ref{the:c1}]
From Lemma \ref{lem:rate}, we know that $I_q$ is
continuous and strictly decreasing from $\ln q$ to $0$
on $(1,\mu_q)$.
Moreover, by Lemma \ref{lem:rate}(i) and the definition
of $\psi_q$, we have
\begin{align*}
I_q(3/2)
&=\frac32\theta_q(3/2)-\psi_q(\theta_q(3/2))=\ln q+
\ln\left(
\frac{e^{3\theta_q(3/2)/2}}
{\sum_{j=1}^q e^{j\theta_q(3/2)}}
\right).
\end{align*}
For $q\geq3$, we obtain from the arithmetic--geometric
mean inequality that
\begin{align*}
\sum_{j=1}^q e^{j\theta_q(3/2)}
&>e^{\theta_q(3/2)}+e^{2\theta_q(3/2)}\geq
2\sqrt{e^{\theta_q(3/2)}e^{2\theta_q(3/2)}}
=2e^{3\theta_q(3/2)/2}.
\end{align*}
Consequently, $I_q(3/2)<\ln q+\ln(1/2)=\ln(q/2)$.
Hence, the unique solution of
$I_q(\rho_q)=\ln(q/2)$ satisfies $1<\rho_q<3/2$.

We choose the reference length so that fewer than
$2^{n-1}$ sequences are excluded.
Let $t_n\triangleq T(L_q(n,T),2^{n-1})$ and
$T_n\triangleq(q+1)n-t_n$.
For $n\geq2$, we have $n<t_n\leq qn$ because
$L_q(n,n)=1<2^{n-1}\leq L_q(n,qn)=q^n$.
From Lemma \ref{lem:finv} with
$R=\log_q2$, $B=1/2$, and $a=0$, we obtain
$t_n=\rho_qn+\ln n/(2\lvert\theta_q(\rho_q)\rvert)+O_q(1)$.
Since $Z_n$ and $(q+1)n-Z_n$ have the same distribution,
the definition of $t_n$ gives
\begin{align}
q^n-L_q(n,T_n)
&=q^n\Prob\{Z_n>T_n\}
=q^n\Prob\{Z_n\leq t_n-1\}=L_q(n,t_n-1)<2^{n-1}.
\label{eq:c1loss}
\end{align}
Furthermore, $q+1-\rho_q-\frac{3q}{4}>\frac{q-2}{4}>0.$
Choose $\delta\triangleq(q+1-\rho_q-3q/4)/2>0$,
which depends only on $q$.
Since $T_n/n\to q+1-\rho_q$, we have
$T_n\geq(3q/4+\delta)n$ for all sufficiently large $n$.
By Lemma \ref{lem:comp} and equation \eqref{eq:c1loss},
there exists a single-deletion correcting code
$\C\subseteq\Sub_n(\A_{q,T_n})$ satisfying
\begin{align*}
\lvert\C\rvert
&\geq
\frac{L_q(n,T_n)+(1-O_q(n^{-1/2}))2^n}{qn}\\
&>\frac{q^n+(1/2-O_q(n^{-1/2}))2^n}{qn}
>\frac{q^n}{qn}
\end{align*}
for all sufficiently large $n$.
Choose $\C'\subseteq\C$ with $\lvert\C'\rvert=M_{n,1}$.
By the definition of $\tau_q^{\del}(n,M_{n,1})$,
we conclude that
\begin{align*}
\tau_q^{\del}(n,M_{n,1})
&\leq T_n
=(q+1-\rho_q)n
-\frac{\ln n}{2\lvert\theta_q(\rho_q)\rvert}
+O_q(1).
\end{align*}
\end{proof}

\section{Conclusion}
\label{sec:con}

We studied the minimum synthesis time for code sizes
$M_n=\lfloor Aq^{Rn}\rfloor$,
$M_n=\lfloor q^n/(nq^{s_n})\rfloor$, and
$M_{n,c}=\lfloor q^n/(nq^c)\rfloor$.
These choices cover a fixed limiting code rate below one,
redundancy $\log_2 n+s_n\log_2q+o(1)$ with $s_n\to\infty$
and $s_n=o(n)$, and redundancy
$\log_2 n+c\log_2q+o(1)$ with fixed $c$, respectively.
We combined finite-length synthesis-time bounds with
probability estimates to determine the minimum time
or bound it in these cases.

For $c=1$, we improved the general upper bound $qn$
by an amount proportional to $n$, but the first-order
minimum synthesis time remains undetermined.
The feasibility of $M_{n,c}$ for
$\log_q(q-1)\leq c<1$ is also not settled by our analysis.
Future work will focus on these cases and on narrowing
the gaps between the synthesis-time bounds.

\appendices
\section{Proof of Lemma \ref{lem:hinv}}
\label{app:hinv}

\begin{lemhinv}
Let $q\geq2$, $a>0$, and $\ell\in\{0,1\}$ be fixed.
Suppose that $s_n\to\infty$ and $s_n=o(n)$.
Let $\gamma_n\in(1,\mu_q)$ be the unique solution of
$I_q(\gamma_n)=s_n\ln q/n$ for all sufficiently large $n$,
and define $\theta_n\triangleq I_q'(\gamma_n)<0$.
If $K_n$ is a positive integer satisfying
$K_n/q^{n-\ell}=aq^{-s_n}(1+o(1))$, then
\begin{align*}
T(L_q(n-\ell,T),K_n)
&=n\gamma_n+
\frac{\frac12\ln(4\pi s_n\ln q)+\ln a}{\lvert\theta_n\rvert}
+o\left(\frac1{\lvert\theta_n\rvert}\right).
\end{align*}
\end{lemhinv}

\begin{proof}
We first consider $\ell=0$ and let
$t_n\triangleq T(L_q(n,T),K_n)$.
By the definition of $t_n$, we have
$L_q(n,t_n-1)<K_n\leq L_q(n,t_n)$.
Since $s_n\to\infty$ and $s_n=o(n)$,
we have $1<K_n<q^n$ for all sufficiently large $n$,
so $n<t_n\leq qn$.
Moreover,
\begin{gather*}
\ln K_n
=n\ln q-s_n\ln q+\ln a+o(1).
\end{gather*}
From equation \eqref{eq:tail-log} and
$L_q(n,T)/q^n=\Prob\{Z_n\leq T\}$,
we know that if $T$ is an integer satisfying
$T/n<\mu_q$ for all sufficiently large $n$,
$T/n\to\mu_q$, and $n(\mu_q-T/n)^2\to\infty$, then
\begin{gather}
\ln L_q(n,T)
=n\ln q-nI_q(T/n)
-\frac12\ln\bigl(4\pi nI_q(T/n)\bigr)+o(1).
\label{eq:hlog}
\end{gather}

We first verify these conditions for $t_n$ and $t_n-1$.
By Lemma \ref{lem:tilt}(i) and Taylor's formula,
as $\theta\to0$, we have
\begin{gather*}
\psi_q(\theta)
=\mu_q\theta+\frac{\sigma_q^2}{2}\theta^2+o(\theta^2),
\quad
\psi_q'(\theta)
=\mu_q+\sigma_q^2\theta+o(\lvert\theta\rvert).
\end{gather*}
For $u\to\mu_q$ with $u<\mu_q$, we have
$\theta_q(u)\to0$.
Together with
$u=\psi_q'(\theta_q(u))$
and Lemma \ref{lem:rate}(i), we obtain
\begin{gather*}
\mu_q-u
=-\sigma_q^2\theta_q(u)+o(\lvert\theta_q(u)\rvert)
=\sigma_q^2\lvert\theta_q(u)\rvert(1+o(1)),\\
I_q(u)
=\theta_q(u)\psi_q'(\theta_q(u))
-\psi_q(\theta_q(u))
=\frac{\sigma_q^2}{2}\theta_q(u)^2(1+o(1)).
\end{gather*}
Consequently,
\begin{gather}
I_q(u)
=\frac{(\mu_q-u)^2}{2\sigma_q^2}(1+o(1)),
\quad
\lvert I_q'(u)\rvert
=\frac{\mu_q-u}{\sigma_q^2}(1+o(1)).
\label{eq:near}
\end{gather}
In particular, since $I_q(\gamma_n)=s_n\ln q/n\to0$,
we have $\gamma_n\to\mu_q$, $\theta_n\to0$, and
$n\sigma_q^2\theta_n^2=2s_n\ln q(1+o(1))$.

To compare $L_q(n,T)$ with $K_n$, for
$c\in\{1/2,2\}$ let $u_n(c)\in(1,\mu_q)$ satisfy
$nI_q(u_n(c))=cs_n\ln q$.
These solutions exist for all sufficiently large $n$
because $cs_n\ln q/n\to0$.
From equation \eqref{eq:near}, we have
$u_n(c)\to\mu_q$ and
$n(\mu_q-u_n(c))^2
=2\sigma_q^2cs_n\ln q(1+o(1))\to\infty$.
Moreover,
\begin{gather*}
\frac{\lvert\lfloor nu_n(c)\rfloor/n-u_n(c)\rvert}
{\mu_q-u_n(c)}
\leq\frac1{n(\mu_q-u_n(c))}
=o(1).
\end{gather*}
Thus, $\lfloor nu_n(c)\rfloor$ also satisfies
the conditions of equation \eqref{eq:hlog}.
By the mean value theorem, $\left\lvert
nI_q\bigl(\lfloor nu_n(c)\rfloor/n\bigr)
-cs_n\ln q
\right\rvert
\leq\lvert I_q'(\xi_{n,c})\rvert
=o(1),$
where $\xi_{n,c}$ lies between
$\lfloor nu_n(c)\rfloor/n$ and $u_n(c)$,
so $\xi_{n,c}\to\mu_q$.
It follows from equation \eqref{eq:hlog}
and the expression for $\ln K_n$ that
\begin{gather*}
\ln\frac{L_q(n,\lfloor nu_n(c)\rfloor)}{K_n}
=(1-c)s_n\ln q
-\frac12\ln(4\pi cs_n\ln q)-\ln a+o(1).
\end{gather*}
Since $\ln(s_n\ln q)=o(s_n\ln q)$,
this expression tends to $-\infty$ for $c=2$
and to $+\infty$ for $c=1/2$.
Therefore, for all sufficiently large $n$,
$L_q(n,\lfloor nu_n(2)\rfloor)
<K_n\leq L_q(n,\lfloor nu_n(1/2)\rfloor)$ and
$\lfloor nu_n(2)\rfloor
<t_n\leq\lfloor nu_n(1/2)\rfloor$.
By the integrality of $t_n$, we have
$u_n(2)<t_n/n\leq u_n(1/2)$.
Hence, $t_n/n<\mu_q$, $t_n/n\to\mu_q$, and
\begin{gather*}
n(\mu_q-t_n/n)^2
\geq n(\mu_q-u_n(1/2))^2\to\infty.
\end{gather*}
The same conditions hold for $t_n-1$, since
$(t_n-1)/n=t_n/n-1/n$.
In addition, the strict decrease of $I_q$ gives $\frac12s_n\ln q
\leq nI_q(t_n/n)<2s_n\ln q.$

We next compare $L_q(n,t_n)$ with $K_n$.
By the mean value theorem, we have $n\bigl(I_q((t_n-1)/n)-I_q(t_n/n)\bigr)
=-I_q'(\xi_n)=o(1),$
where $\xi_n$ lies between $(t_n-1)/n$ and $t_n/n$.
The last equality follows from $\xi_n\to\mu_q$
and $I_q'(\xi_n)\to0$.
Since $nI_q(t_n/n)\to\infty$, we also obtain
\begin{gather*}
\ln\frac{I_q((t_n-1)/n)}{I_q(t_n/n)}
=\ln\left(1+\frac{o(1)}{nI_q(t_n/n)}\right)
=o(1).
\end{gather*}
Then, from equation \eqref{eq:hlog},
\begin{align*}
0
&\leq\ln L_q(n,t_n)-\ln L_q(n,t_n-1)\\
&=n\bigl(I_q((t_n-1)/n)-I_q(t_n/n)\bigr)
+\frac12\ln\frac{I_q((t_n-1)/n)}{I_q(t_n/n)}
+o(1)=o(1).
\end{align*}
Together with $L_q(n,t_n-1)<K_n\leq L_q(n,t_n)$,
we have
\begin{gather*}
0\leq\ln L_q(n,t_n)-\ln K_n
<\ln L_q(n,t_n)-\ln L_q(n,t_n-1)
=o(1).
\end{gather*}
Thus, $\ln L_q(n,t_n)=\ln K_n+o(1)$.
Using equation \eqref{eq:hlog} and
$\ln K_n=n\ln q-s_n\ln q+\ln a+o(1)$,
we obtain
\begin{gather}
nI_q(t_n/n)
+\frac12\ln\bigl(4\pi nI_q(t_n/n)\bigr)
=s_n\ln q-\ln a+o(1).
\label{eq:hbal}
\end{gather}
Since $\frac12s_n\ln q\leq nI_q(t_n/n)<2s_n\ln q$,
we have
$\ln(4\pi nI_q(t_n/n))=O(\ln(s_n\ln q))$.
It follows from equation \eqref{eq:hbal} that
\begin{gather*}
nI_q(t_n/n)=s_n\ln q+O(\ln(s_n\ln q)),\\
\ln\frac{nI_q(t_n/n)}{s_n\ln q}
=\ln\left(
1+O\left(\frac{\ln(s_n\ln q)}{s_n\ln q}\right)
\right)
=o(1).
\end{gather*}
Consequently,
$\ln(4\pi nI_q(t_n/n))
=\ln(4\pi s_n\ln q)+o(1)$.
Substitution into equation \eqref{eq:hbal} gives
\begin{gather}
nI_q(t_n/n)
=s_n\ln q-\frac12\ln(4\pi s_n\ln q)-\ln a+o(1).
\label{eq:hlev}
\end{gather}

We now determine
$\Delta_n\triangleq t_n-n\gamma_n$.
From equations \eqref{eq:near} and \eqref{eq:hlev},
together with $nI_q(\gamma_n)=s_n\ln q$, we have
\begin{gather*}
\frac{\mu_q-t_n/n}{\mu_q-\gamma_n}
=
\sqrt{\frac{I_q(t_n/n)}{I_q(\gamma_n)}}(1+o(1))
=1+o(1).
\end{gather*}
By the mean value theorem, $n\bigl(I_q(t_n/n)-I_q(\gamma_n)\bigr)
=I_q'(\eta_n)\Delta_n,$
where $\eta_n$ lies between $\gamma_n$ and $t_n/n$.
Therefore,
\begin{gather*}
\min\left\{1,\frac{\mu_q-t_n/n}{\mu_q-\gamma_n}\right\}
\leq\frac{\mu_q-\eta_n}{\mu_q-\gamma_n}
\leq
\max\left\{1,\frac{\mu_q-t_n/n}{\mu_q-\gamma_n}\right\},
\end{gather*}
so $(\mu_q-\eta_n)/(\mu_q-\gamma_n)\to1$.
Since $I_q'(\eta_n)<0$ and $\theta_n=I_q'(\gamma_n)<0$,
equation \eqref{eq:near} implies
\begin{gather*}
\frac{I_q'(\eta_n)}{\theta_n}
=\frac{\mu_q-\eta_n}{\mu_q-\gamma_n}(1+o(1))
=1+o(1).
\end{gather*}
Together with equation \eqref{eq:hlev}, we obtain $I_q'(\eta_n)\Delta_n
=-\frac12\ln(4\pi s_n\ln q)-\ln a+o(1),$
which gives
$\Delta_n=O(\ln(s_n\ln q)/\lvert\theta_n\rvert)$.
Using $n\sigma_q^2\theta_n^2=2s_n\ln q(1+o(1))$,
we have
\begin{gather*}
\frac{\Delta_n^2}{n}
=O\left(
\frac{\ln^2(s_n\ln q)}{n\theta_n^2}
\right)
=O\left(
\frac{\ln^2(s_n\ln q)}{s_n\ln q}
\right)
=o(1).
\end{gather*}
Moreover, from Lemma \ref{lem:rate}(ii),
we know that $I_q''$ is bounded in a left
neighborhood of $\mu_q$.
By Taylor's formula and $I_q'(\gamma_n)=\theta_n$,
\begin{gather*}
n\bigl(I_q(t_n/n)-I_q(\gamma_n)\bigr)
=
\theta_n\Delta_n
+\frac12 I_q''(\zeta_n)\frac{\Delta_n^2}{n}
=
\theta_n\Delta_n+o(1),
\end{gather*}
where $\zeta_n$ lies between $\gamma_n$ and $t_n/n$.
The last equality follows from
$I_q''(\zeta_n)=O_q(1)$ and $\Delta_n^2/n=o(1)$.
Combining this identity with equation \eqref{eq:hlev}
and $nI_q(\gamma_n)=s_n\ln q$, we obtain $\theta_n\Delta_n
=-\frac12\ln(4\pi s_n\ln q)-\ln a+o(1).$
Since $\theta_n<0$, we conclude that $t_n=n\gamma_n+\Delta_n
=n\gamma_n+
\frac{\frac12\ln(4\pi s_n\ln q)+\ln a}
{\lvert\theta_n\rvert}
+o\left(\frac1{\lvert\theta_n\rvert}\right).$
This proves the result for $\ell=0$.

For $\ell=1$, we apply the proved case to $qK_n$,
since
$qK_n/q^n=K_n/q^{n-1}=a q^{-s_n}(1+o(1))$.
Moreover, $1\leq K_n\leq q^{n-1}$ for all sufficiently
large $n$.
By Lemma \ref{lem:shift}, we have $T(L_q(n-1,T),K_n)
=T(L_q(n,T),qK_n)+O_q(1).$
As $\theta_n\to0$, the additional $O_q(1)$ term
is $o(1/\lvert\theta_n\rvert)$.
Hence, the same expansion holds for $\ell=1$,
which completes the proof.
\end{proof}

\bibliographystyle{IEEEtran}
\bibliography{reference}
\end{document}